\documentclass[aps,prd,nofootinbib,superscriptaddress,preprint,eqsecnum,showkeys,showpacs,preprintnumbers]{revtex4-2}
\usepackage{slashed}
\usepackage{graphics}
\usepackage{graphicx}
\usepackage{subcaption}
\usepackage{dcolumn}
\usepackage{epsfig}
\usepackage{mathtools} %以下几个是数学包
\usepackage{amsmath,bm,amsfonts} %数学字体包等
\usepackage{amsthm,amssymb}
\usepackage{mathrsfs}
\usepackage{ulem} %加下划线,文字划掉效果,配合"\sout{文字}"
\usepackage{cancel} %文字消除效果，"\bcancel{},\cancel{},\xcance{},\cancelto{}{}"
\usepackage{marginnote} %加边注
\usepackage{pifont} %带圈序号
\usepackage{listings} %加代码包
\usepackage{framed} %加边框
\usepackage{appendix} %加附录
\usepackage{color} %加颜色
\usepackage[greek,english]{babel}%可以编译希腊字母等
\usepackage{xtab} %制作线表工具
\usepackage{multirow,booktabs}%三线表和表格控制
\usepackage{shapepar}%文字绕排
\usepackage{tikz-cd,lipsum} %绘图包
\usetikzlibrary{positioning, shapes.geometric}
\usepackage{tkz-euclide}
\usepackage[colorlinks=true,linkcolor=blue,citecolor=blue,urlcolor=blue]{hyperref} 
\usepackage{orcidlink}
\allowdisplaybreaks[3]

\definecolor{tianlan}{RGB}{102,179,255}
\newcommand{\circledtext}[2]{%
	\protect\tikz[baseline=(char.base)]{%
		\protect\node[shape=circle,draw=tianlan,fill=#1,inner sep=2pt, text=white] (char) {#2};%
	}%
}

\begin{document}

\title{Energy flux and waveforms by coalescing  spinless binary system in effective one-body theory}

\author{Sheng Long\,\orcidlink{0009-0009-0163-2724}}
%\email[ ]{643351248@qq.com}
 \affiliation{Department
of Physics, Key Laboratory of Low Dimensional Quantum Structures and
Quantum Control of Ministry of Education, and Synergetic Innovation
Center for Quantum Effects and Applications, Hunan Normal
University, Changsha, Hunan 410081, P. R. China}

\author{Weike Deng\footnote{Weike Deng and Sheng Long contributed equally to this work and should be considered as co-first author.}}
%\email[ ]{1721458793@qq.com}
\affiliation{Department
	of Physics, Key Laboratory of Low Dimensional Quantum Structures and
	Quantum Control of Ministry of Education, and Synergetic Innovation
	Center for Quantum Effects and Applications, Hunan Normal
	University, Changsha, Hunan 410081, P. R. China}
	
\author{{Jiliang} {Jing}\,\orcidlink{0000-0002-2803-7900}, } 
\email[Corresponding author: ]{jljing@hunnu.edu.cn}
\affiliation{Department of Physics, Key Laboratory of Low Dimensional Quantum Structures and Quantum Control of Ministry of Education, and Synergetic Innovation
Center for Quantum Effects and Applications, Hunan Normal
University, Changsha, Hunan 410081, P. R. China}
\affiliation{Center for Gravitation and Cosmology, College of Physical Science and Technology, Yangzhou University, Yangzhou 225009, P. R. China}

%\date{\today}

\begin{abstract}
We present a study on the energy radiation rate and waveforms of the gravitational wave generated by coalescing spinless binary systems up to the third post-Minkowskian approximation in the effective one-body theory. To derive an analytical expansion of the null tetrad components of the gravitational perturbed Weyl tensor $\varPsi_{4}$ in the effective spacetime, we utilize the method proposed by Sasaki $et$ $al.$ During this investigation, we discover more general integral formulas that provide a theoretical framework for computing the results in any order. Subsequently, we successfully compute the energy radiation rate and waveforms of the gravitational wave, which include the results of the Schwarzschild case and the correction terms resulting from the dimensionless parameters $a_{2}$ and $a_{3}$ in the effective metric.
\end{abstract}

\maketitle
\newpage

\section{Introduction}
Gravitational waveform templates play an important role in the detection of gravitational wave events generated by coalescing binary systems \cite{Abbott_2016_physrevlett.116.241103,PhysRevLett.118.221101,Abbott_2017_physrevlett.119.141101,Abbott_2017_physrevlett.119.161101,Abbott_2020_physrevd.102.043015,Abbott_2021_physrevx.11.021053,Abbott_2022_physrevd.105.122001,Nitz_2021_ac1c03,LIGOScientific:2021djp,LIGOScientific:2021usb}. The foundation of gravitational waveform templates is the theoretical model of gravitational radiation, in which the key point is studying the late-stage dynamical evolution of a coalescing binary system.

Damour and Buonanno \cite{Damour_1988_10.1007/bf02828697,Buonanno_1999_physrevd.59.084006} proposed an effective one-body (EOB) theory that maps the real two-body problem with masses $m_{1}$ and $m_{2}$ to a test particle of mass $\mu = \frac{m_{1} m_{2}}{m_{1} + m_{2}}$ moving around an effective spacetime of mass $M = m_{1} + m_{2}$ (and we denote the symmetric mass ratio as $\nu = \mu / M$). This theory enables the study of gravitational radiation produced by merging binary systems. Based on the EOB theory with the post-Newtonian (PN) approximation, Damour $et$ $al.$ provided an estimate of the gravitational waveforms emitted throughout the inspiral, plunge, and coalescence phases \cite{Buonanno_2007_physrevd.75.124018,Damour_2000_physrevd.62.084011}.

To release the assumption that $v/c$ is a small quantity, in 2016, Damour introduced another theoretical model by combining the EOB theory with the post-Minkowskian (PM) approximation \cite{Damour_2016_physrevd.94.104015,Bini_2018_physrevd.98.044036}. Damour and Rettegno \cite{damour2023strongfield} compared numerical relativistic (NR) data for equal-mass binary black hole scattering with analytical predictions based on the fourth PM (4PM) dynamics \cite{Dlapa_2022_physrevlett.128.161104, Dlapa_2023_physrevlett.130.101401, K_lin_2020_physrevlett.125.261103, DAMOUR_2008_s0217751x08039992, Damour_2008_physrevd.77.024043, damour2009effective, Damour_2007_physrevd.76.064028, Damour:2019lcq} and pointed out that the reconstruction of PM information in terms of EOB radial potentials leads to remarkable agreement with NR data, especially when using radiation-reacted 4PM information. Therefore, this new model may lead to a theoretically improved version of the EOB conservative dynamics and may be useful in the upcoming era of high-signal-to-noise-ratio gravitational wave observations.

The dynamical evolution of a coalescing binary system for a spinless EOB theory can be described by the Hamilton equation \cite{Buonanno_2000_physrevd.62.064015}, and the Hamiltonian $H[g_{\mu \nu}^{eff}]$ is dependent on the effective metric. The radiation reaction forces $\mathcal{F}_{R}[g_{\mu \nu}^{eff}]$ and $\mathcal{F}_{\varphi}[g_{\mu \nu}^{eff}]$ in the Hamilton equation can be described by the energy radiation rate as follows: $\frac{\mathrm{d} E}{\mathrm{d} t} = \frac{1}{4 \pi G^{2} \omega^{2}} \int |\varPsi_{4}^{B}|^{2} r^{2} \mathrm{d} \Omega$ \cite{Poisson_1993_physrevd.47.1497,Buonanno_2007_physrevd.75.124018}. Furthermore, the “plus” and “cross” modes of gravitational waves are related to the null tetrad components of the gravitational perturbed Weyl tensor $\varPsi_{4}^{B}$ in the Newman--Penrose formalism as follows: $\varPsi_{4}^{B} = \frac{1}{2} (\ddot{h}_{+} - i \ddot{h}_{\times})$. Thus, as long as we obtain the effective metric and the solution of $\varPsi_{4}^{B}$ in the Newman--Penrose formalism, we can calculate the energy radiation rate and construct gravitational waveforms.

In previous work, we attempted to develop a self-consistent EOB theory for spinless and spinning binaries based on the PM approximation \cite{Long_2023,Jing:2022vks,Jing_2023_s11433-023-2084-1,2021EPJC...81...97H}. Furthermore, in a recent paper \cite{Jing_2023_4PM_metric}, we obtained the effective metric up to the 4PM order. We adopted the black hole perturbation method used by Teukolsky \cite{1973ApJ_185_635T,1973JMP_14_7B} and decomposed all quantities into background and perturbation (denoted with a superscript $B$) parts in the Newman–Penrose formalism. After choosing a shadow gauge \cite{Chandrasekhar1975OnTE,1983mtbh.book_C,Jing:2022vks} with $\varPsi_{1}$ and $\varPsi_{3}$ set to $0$, we can decouple the equations for the null tetrad components of the gravitational perturbed Weyl tensor $\varPsi_{4}^{B}$. Subsequently, upon separating the variables in the equations, we obtained a radial equation, which is the so-called Teukolsky-like equation, and an angular equation that features spin-weighted spherical harmonics.

This Teukolsky-like equation is more complex compared to the Teukolsky equation in Kerr and Schwarzschild spacetimes. We were unable to find a similar transformation to convert the homogeneous Teukolsky-like equation into hypergeometric or Heun equations; thus, we did not choose to adopt the so-called MST \cite{Mano_1996,Sasaki_2003} method or the Heun function \cite{Cook_2014,Fiziev_2011,1986JMP_27.1238L,1985_Le,Leaver_1986,chen2023radiation}. Instead, we follow the approach used by Sasaki. Several researchers \cite{Han_2011, Nagar_2017_1111, Nagar_2022_111, PhysRevD.95.044028, Xin:2021zir, Zhang:2021fgy, Zhang:2020rxy, Cheng:2017xwm, Cao:2017ndf, Gamba:2021ydi, Ossokine:2020kjp, Damour:2014sva, LISAConsortiumWaveformWorkingGroup:2023arg} have employed numerical methods to solve the Teukolsky equation and achieved significant success by combining the EOB theory with numerical relativity.

We initially applied the Sasaki--Nakamura--Chandrasekhar-like (S-N-C-like) transformation \cite{Chandrasekhar1975OnTE,Sasaki_1994} to convert the homogeneous Teukolsky-like equation into a homogeneous Sasaki--Nakamura-like (S-N-like) equation \cite{Poisson_1993,Sasaki_1994,Mino_1997,Sasaki_2003}. In an asymptotically flat spacetime, this homogeneous S-N-like equation can be simplified to the Klein–Gordon equation. Subsequently, we performed a Taylor expansion with respect to $\eta = 2GM\omega$. The equation of the zeroth order is the spherical Bessel equation, and its solutions are linear combinations of the first and second kind spherical Bessel functions, denoted as $j_\ell$ and $n_\ell$, respectively, allowing us to construct higher-order solutions based on the zeroth-order solution. By performing an inverse transformation, we can deduce the solutions of the homogeneous Teukolsky-like equation. This framework enables us to construct the solutions of the inhomogeneous Teukolsky equation, which includes a source term, using Green’s function method.

However, due to the complexity of Green’s function method and the fact that the integral formulas provided in the previous work of Sasaki \cite{Mino_1997} were insufficient for our needs in computing higher-order solutions, we found some new integral formulas presented in Appendix \ref{ap:general} of this article, which are crucial for our journey toward calculating higher-order solutions.

Section II introduces the effective metric of 3PM, while Sec. III discusses the solutions of the equation for $\varPsi^B_4$ in the effective spacetime. Specifically, Section IIIA summarizes the general structure of the solutions of the radial equation (Teukolsky-like equation) of $\varPsi_{4}^{B}$. Section IIIB provides a comprehensive explanation of the calculation of the homogeneous S-N-like equation. Subsequently, we employ boundary conditions to determine the amplitudes. Section IIIC presents the source terms for quasi-circular orbits; by combining the homogeneous solutions provided in Section IIIB and utilizing Eq.~\eqref{ZZSch}, we can obtain the solution of $\varPsi_{4}^{B}$ under quasi-circular orbits. In Section IV, we present the energy radiation rate $\frac{\mathrm{d} E}{\mathrm{d} t}$ and the gravitational waveforms $h_{\ell m}$.

\section{Effective metric for the EOB theory }\label{sec:EOBM}
In the EOB theory, the main idea is to map the two-body problem onto an EOB problem, that is, a test particle orbits around a massive black hole described by an effective metric. With the help of the scattering angles, we found that the effective metric for spinless binaries with radiation reaction effects in the EOB theory, up to the 3PM approximation, can be expressed as follows \cite{Jing_2023_4PM_metric}:
\begin{eqnarray}
ds_{\rm eff}^2=g^{\text{eff}}_{\mu\nu}d x^\mu d x^\nu=\frac{\Delta_{r}}{r^2} dt^2-\frac{r^2}{\Delta_{r}}dr^2- r^2(d\theta^2+\sin^2\theta d\varphi^2),\label{Mmetric}
\end{eqnarray}
with
\begin{align}\label{Delta}
\nonumber \Delta_{r} = & r^2- 2 GM r+a_2 (GM)^2+ a_3\frac{ (GM)^3}{r} \\
           = & \frac{1}{r}(r-2c_h GM)(r-2c_1 GM)(r-2c_2 GM),
\end{align}
the definitions of $c_{1}, c_{2},$ and $c_{h}$ are as follows:
\begin{align}
	c_{1} = & \frac{1}{3} - \frac{1}{2} \Big[ (1 - i \sqrt{3}) (Q + \sqrt{P^{3} + Q^{2}})^{\frac{1}{3}} + (1 + i \sqrt{3}) (Q - \sqrt{P^{3} + Q^{2}})^{\frac{1}{3}} \Big] , \\
	c_{2} = & \frac{1}{3} - \frac{1}{2} \Big[ (1 + i \sqrt{3}) (Q + \sqrt{P^{3} + Q^{2}})^{\frac{1}{3}} + (1 - i \sqrt{3}) (Q - \sqrt{P^{3} + Q^{2}})^{\frac{1}{3}} \Big] , \\
	c_{h} = & \frac{1}{3} + (Q + \sqrt{P^{3} + Q^{2}})^{\frac{1}{3}} + (Q - \sqrt{P^{3} + Q^{2}})^{\frac{1}{3}} ,
\end{align}
where $Q = \frac{1}{27} - \frac{a_{2}}{24} - \frac{a_{3}}{16}$, $P = \frac{1}{3} \Big( \frac{a_{2}}{4} - \frac{1}{3} \Big)$ and $a_{2}$ and $a_{3}$ are dimensionless parameters expressed as follows:
\begin{align}
& a_2 = \frac{3(1- \, \Gamma)(1-5 \, \gamma^2)}{\Gamma\, (3\, \gamma^2-1 )} , \\
&a_3 = \frac{3}{2 (4 \gamma^{2} - 1)}  \Big[\frac{3- 2\, \Gamma-3 (15 -8\, \Gamma ) \gamma^2+6 ( 25-16 \, \Gamma ) \gamma^4}{\Gamma\, (3\, \gamma^2-1 )} - 2 P_{30} - \frac{2 \chi_{3}^{rr}}{ \sqrt{\gamma^{2} - 1}}\Big],\label{a3}
\end{align}
in which $\gamma = \frac{E}{\mu} = \frac{1}{2} \frac{\mathcal{E}^2-m_1^2-m_2^2}{m_1m_2} $ is the Lorentz factor variable, $\mathcal{E}$ is the real two-body energy \cite{Jing_2023_4PM_metric,Damour_2023}, $E$ is the effective energy, $\Gamma =  E/M = \sqrt{1+2 \nu (\gamma - 1)}$ is the rescaled energy, and
\begin{align}
\nonumber P_{30} = & \frac{18\gamma^2-1}{2\,\Gamma^2}+\frac{8\,\nu\, (3+12\gamma^2-4\gamma^4)}{\Gamma^2\, \sqrt{\gamma^2-1}} \mbox{arcsinh}\sqrt{\frac{\gamma-1}{2}}
+ \frac{\nu}{\Gamma^2} \big( 1 - \frac{103}{3}\gamma - 48 \gamma^2 - \frac{2}{3} \gamma^3 + \\
& \frac{3 \,\Gamma\,(1-2\gamma^2)(1-5 \gamma^2)}{(1+\Gamma)(1+\gamma)} \big) , \\
\chi_{3}^{rr} = & - \frac{2 \nu }{\Gamma^{2}} \frac{\gamma (2 \gamma^{2} - 1)^{2}}{3 (\gamma^{2} - 1)^{3/2}} \bigg\{ \frac{(5 \gamma^{2} - 8)}{\gamma} \sqrt{\gamma^{2} - 1} + 2 (9 - 6 \gamma^{2}) \mbox{arcsinh} \sqrt{\frac{\gamma-1}{2}} \bigg\} .\label{x3rr}
\end{align}
In Eq. (\ref{a3}), the term $\chi^{rr}_3$, described by Eq. (\ref{x3rr}), represents the 3PM radiation reaction effects, which shows that the structure of the effective spacetime is affected by the radiation reaction effect.

\section{Solutions of equation for $\varPsi^B_4$ in effective spacetime}\label{sec:slepsi4}
In this section, we first present the formal solution of the radial equation for $\varPsi^B_4$. Then, we transform the radial equation without source to the corresponding S-N-like equation, and we look for its solution. At last, we work out the solution of the radial equation of $\varPsi^B_4$ with the source, which describes the gravitational radiation induced by the motion of an effective particle in an effective background.

\subsection{Formal solution of the radial equation of $\varPsi_4^B$ in effective spacetime} \label{subsec:formalsolution}
In the EOB theory for the spinless real two-body system, we have found a decoupled equation of $\varPsi^B_{4}$ for the gravitational perturbation in the effective spacetime \eqref{Mmetric} using the gauge transform property of the tetrad components of the perturbed Weyl tensors and separated the decoupled equation in the radial and angular parts, in which the radial part of $\varPsi_4^B$ is given by \cite{Jing:2022vks,Long_2023}
\begin{eqnarray} \label{EqTS}
\left[ \frac{\Delta_{r}^{2}}{\mathfrak{f}}  \frac{\mathrm{d} }{\mathrm{d} r}  \Big(\frac{\mathfrak{f}}{\Delta_{r}} \frac{\mathrm{d} }{\mathrm{d} r}  \Big) - V (r) \right] R _{lm\omega}(r)= T _{\ell m \omega}(r),
\end{eqnarray}
with
\begin{align}
&\mathfrak{f} = -  \frac{3 G M}{r^{3} F_{4}}, \nonumber  \\  &V (r) =  -\frac{r^{2} \omega(r^{2} \omega+2 i \Delta_{r}') }{\Delta_{r}} + i r \omega \big(5  +\frac{2  r^{3}  F_1  }{\Delta_{r}} \big) - \frac{3 \Delta_{r}}{r^{2}} - \frac{3 {\Delta_{r}'}}{r} + 6 r F_{1} - 2 r^2 F_{4} + \lambda F_{2},   \nonumber  \\
&T _{\ell m\omega}(r)=-\mu \int^{\infty}_{-\infty}dt
e^{i\omega t-i m \varphi(t)}
\Delta_{r}^2\Big\{ \mathbf{A}_{0} \delta(r-r(t)) + \Big[ \mathbf{A}_{1}
\delta(r-r(t))\Big]^{'}+\Big[\mathbf{A}_{2} \delta(r-r(t))\Big]^{''}\Big\},
\label{TgenTTsl}
\end{align}
where the prime $'$ denotes derivation with respect to $r$, and $\mathbf{A}_{0} = A _{n n\,0}+A _{{\overline m} n\,0}
+A _{{\overline m} {\overline m}\,0}$, $\mathbf{A}_{1} = A _{{\overline m} n\,1}
+A _{{\overline m} {\overline m}\,1}$, and $\mathbf{A}_{2} = A_{{\overline m} {\overline m}\,2}$
\begin{align}
	& \mathbf{A} _{nn\,0} = -\frac{2\,r^4 }{ \sqrt{2\pi}\,\Delta_{r}^2}\,
	C _{n n}  \,F_2 \,
	\mathscr{L}_1^+\Big[\mathscr{L}_2^+\Big(\ _{-2}Y_{\ell m}(\theta)\Big)  \Big],\nonumber \\
	& \mathbf{A} _{{\overline m}n\,0} = \frac{r^3 }{ \sqrt{\pi}\Delta_{r}}
	\,  C _{{\overline m} n} \Big[\big(1+F_2\big)\frac{ i r^2\omega }{ \Delta_{r}}+\frac{F_2}{r} -F_{2}^\prime-\frac{F_4^\prime}{F_4}\Big] \mathscr{L}_2^{\dag} \Big(\ _{-2}Y_{\ell m}(\theta) \Big),\nonumber \\
	& \mathbf{A} _{{\overline m}{\overline m}\,0}
	= \frac{r^2 }{ \sqrt{2\pi}}\,
	C _{{\overline m} {\overline m}}\ _{-2}Y_{\ell m}(\theta)  \Bigl[
	i\Bigl(\frac{ r^2\omega }{ \Delta_{r}}\Bigr)^\prime+\frac{r^4\omega^2 }{ \Delta_{r}^2}+ \frac{ i r^2\omega }{ \Delta_{r}}\Big(\frac{1}{r}+\frac{F_4^\prime}{F_4}\Big)+\frac{F_4^\prime}{r F_4}+\Big(\frac{F_4^\prime}{F_4}\Big)^\prime\Bigr],\nonumber \\
	& \mathbf{A} _{{\overline m}n\,1} = \frac{
		r^3}{ \sqrt{\pi}\Delta_{r} }\,C _{{\overline m} n}
	\big(1+F_2\big) \mathscr{L}_2^{\dag}\Big(\ _{-2}Y_{\ell m}(\theta) \Big)
	,\nonumber \\
	& \mathbf{A} _{{\overline m}{\overline m}\,1}
	= \frac{ r^2 }{ \sqrt{2\pi}}
	\,
	C _{{\overline m}{\overline m}}\ _{-2}Y_{\ell m}(\theta)
	\Bigl( -2 i\frac{ r^2\omega}{ \Delta_{r}}+\frac{1}{r}+\frac{F_4^\prime}{F_4}\Bigr),\nonumber \\
	& \mathbf{A} _{{\overline m}{\overline m}\,2}
	= -\frac{r^2}{ \sqrt{2\pi}}\,
	C _{{\overline m} {\overline m}}\ _{-2}Y_{\ell m}(\theta) ,
	\label{Aijs}
\end{align}
the explicitly definitions of $F_{a} $ $(a = 1, 2, 3, 4)$, $\mathscr{L}_n$ (\text{or}  $\mathscr{L}_n^{\dagger}$), and $C_{b}$ ($b$ is  $nn$ or ${\overline m}n$ or ${\overline m}{\overline m}$ ) can be found in Ref.~\cite{Jing:2022vks}, and $_{-2}Y_{\ell m}(\theta)$ is the spin-weighted spherical harmonics \cite{Sasaki_2003,Dray1985TheRB}.

The radial equation, i.e., Eq. (\ref{EqTS}), can be solved using Green’s function method. That is, based on the homogeneous solutions of Eq.~\eqref{EqTS}\begin{eqnarray}
R^{\rm in}_{\ell m\omega}
\to \left\{\begin{array}{cc}
B^{\rm trans}_{\ell m\omega}\Delta_r^2  e^{-i \omega r^*},
\ \ \ \ \ \ \ \ \ \ \ \ \ \ \ \ \ \ \ \ \ \ \ \ \  & \text{for} \ \ r\to r_+, \ \ \ \  \\
r^3 B^{\rm  ref}_{\ell m\omega}e^{i\omega r^*}+
r^{-1}B^{\rm in}_{\ell m\omega} e^{-i\omega r^*},
\ \ \ \ \ \ & \text{for} \ \  r\to +\infty,
\end{array}\right.
\label{Kk}
\end{eqnarray}
\begin{eqnarray}
R^{\rm up}_{\ell m\omega}
\to \left\{\begin{array}{cc}
C^{\rm  up}_{\ell m\omega} e^{i \omega r^*}+
\Delta_r^2 C^{\rm ref}_{\ell m\omega} e^{-i\omega r^*},
\ \ \ \ \   & \text{for} \ \ r\to r_+, \ \ \ \  \\
C^{\rm trans}_{\ell m\omega} r^3  e^{i\omega r^*},
\ \ \ \ \ \ \ \ \ \ \ \ \ \ \ \ \ \ \ \ \ \ \  & \text{for} \ \  r\to +\infty,
\end{array}\right.
\label{Kkk}
\end{eqnarray}
where $r^*$ denotes the tortoise coordinate defined by $r^{*}=\int \frac{r^2}{ \Delta_r}dr$. The inhomogeneous solution of the radial Eq. \eqref{EqTS} can be expressed as follows:
\begin{align}\label{jx}
	R_{\ell m\omega}=&\frac{1}{2 i \omega C^{\rm trans}_{\ell m\omega}
		B^{\rm inc}_{\ell m\omega}}
	\Big\{R^{\rm up}_{\ell m\omega}\int^r_{r_+}d\tilde{r}
	\frac{ \mathfrak{f} R^{\rm in}_{\ell m\omega}(\tilde{r} )
		T_{\ell m\omega}(\tilde{r} ) }{\Delta_r^{2} (\tilde{r} ) }
	+ R^{\rm in}_{\ell m\omega}\int^\infty_{r}d\tilde{r}
	\frac{ \mathfrak{f} R^{\rm up}_{\ell m\omega}(\tilde{r} )  T_{\ell m\omega}(\tilde{r} ) }{\Delta_r^{2}(\tilde{r} ) }\Big\},
\end{align}
whereas the counterpart at infinity can be expressed as follows:
\begin{equation}
R_{\ell m \omega}(r \rightarrow \infty) \rightarrow \frac{r^3 \mathrm{e}^{\mathrm{i} \omega r^*}}{2 \mathrm{i} \omega B_{\ell m \omega}^{\mathrm{inc}}} \int_{r_{+}}^{\infty} \mathrm{d} \tilde{r} \frac{T_{\ell m \omega}(\tilde{r}) R_{\ell m \omega}^{\mathrm{in}}(\tilde{r})}{\tilde{r}^3 F_4(\tilde{r}) \Delta_r^2(\tilde{r})} \equiv \hat{Z}_{\ell m \omega} r^3 \mathrm{e}^{\mathrm{i} \omega r^*} .
\end{equation}
As discussed in Ref.~\cite{Jing:2022vks}, for the point source case, after a lengthy calculation, we can obtain the expression for $\hat{Z}_{\ell m \omega}$. If we focus our attention just on the quasi-circular orbit, we have $\hat{Z}_{\ell m\omega_n} = Z_{\ell m\omega} \, \delta(\omega - \omega_{n})$, in which
\begin{eqnarray}
 Z _{\ell m\omega}&=&
\frac{\pi \nu G M}{ i\omega B^{\rm inc}_{\ell m\omega}}
\Bigl[ \mathbf{A}_{0}
\mathfrak{f} R_{\ell m \omega}^{in} - \mathbf{A}_{1}
 \frac{d  }{ dr}\Big(\mathfrak{f} R_{\ell m \omega}^{in}\Big) +  \mathbf{A}_{2} \frac{d^2  }{ dr^2}\Big(\mathfrak{f} R_{\ell m \omega}^{in}\Big) \Bigr].
\label{ZZSch}
\end{eqnarray}
Then, the solution of $\varPsi_4^B$ is described by
\begin{equation}
\varPsi^B_4=\frac{1}{ \mathcal{R}}\sum_{\ell m n}
\hat{Z}_{\ell m\omega_n} \frac{{}_{-2}Y_{\ell m} }{ \sqrt{2\pi}}
e^{i\omega_n(r^*-t)+im\varphi}.
\label{psi41}
\end{equation}
Equations \eqref{psi41} and \eqref{ZZSch} show that to get the explicit expression for $\varPsi^B_4$, we should work out $\mathbf{A}_{i} (i = 1, 2, 3)$, $B^{\rm inc}_{\ell m\omega}$, and $R^{\rm in}_{\ell m\omega}$.

\subsection{S-N-like equation and its solution of the third PM approximation} \label{subsec:S-N-likeequation}
In Eq.~\eqref{ZZSch}, $R^{\rm in}_{\ell m\omega}$ is the solution of the homogeneous equation without the source term. To get the solution, we do not treat the Teukolsky-like equation directly because the potential function in the equation is a long-range potential. Instead, we transform the radial equation, i.e., Eq. (ref{EqTS}), without source into the S-N-like equation with a new function $X_{\ell m\omega}$, which has a short-range potential. Then, using the solution of $X^{\rm in}_{\ell m\omega}$, we find out $B^{\rm in}_{\ell m\omega}$ and $R^{\rm in}_{\ell m\omega}$.

\subsubsection{S-N-like equation and relation between $R^{\rm in}_{\ell m\omega}$ and $X^{\rm in}_{\ell m\omega}$}
Taking an S-N-C-like transformation as\footnote{Because the physical quantities we are concerned with are related to the $\ell$, $m$, and $\omega$, for any $\ell$, $\omega$ depends on $m$ and $m = - \ell, - \ell + 1, \dots, 0, \dots, \ell - 1, \ell$. Thus, we shall henceforth represent the subscript labels $\ell m \omega$ of these physical quantities simply as $\ell$ or drop the subscript for the sake of brevity and clarity.}
\begin{eqnarray}\label{ChiR}
X(r) = \frac{r f^{1/2}}{\Delta_{r}} \left( \alpha(r) R(r) + \frac{\beta(r)}{\Delta_{r}} {R}'(r) \right),
\end{eqnarray}
where
\begin{align}
	\alpha(r) = & 6 \frac{\Delta_{r}}{r^{2}} + V(r) + 2 i r \omega + i r^{2} \omega \frac{{\Delta_{r}}'}{\Delta_{r}} - \frac{r^{4} \omega^{2}}{\Delta_{r}} , \nonumber \\
	\beta(r)  = & \Delta_{r} (- 2 i r^{2} \omega - 4 \frac{\Delta_{r}}{r} - \Delta_{r} \frac{{\mathfrak{f}}'}{\mathfrak{f}} + \Delta_{r}) .
\end{align}
and considering the coordinate transformation $r \rightarrow r_{*}$, the radial equation (Eq. \eqref{EqTS}) without the source term ($T _{\ell m \omega} = 0$) can be rewritten as the so-called S-N-like equation, as follows:
\begin{equation} \label{eqX}
{X}_{,r^{*} r^{*}} - \mathscr{F} {X}_{,r^{*}} - \mathscr{U} X = 0 ,
\end{equation}
with
\begin{align}\label{eq:The expression of coefficient mathscrF and mathscrU}
\mathscr{F} & = \frac{\Delta_{r}}{r^{2}} \frac{{\gamma}'}{\gamma} , \nonumber \\
\mathscr{U} & = \frac{\Delta_{r}}{r^{4}} U + G^{2} + G_{,r^{*}} - \frac{\Delta_{r}}{r^{2}} \frac{{\gamma}'}{\gamma} G ,
\end{align}
where
\begin{align}
\gamma & = \alpha \Big( \alpha + \frac{{\beta }'}{\Delta_{r}} - \frac{\beta}{\Delta_{r}} \frac{ {\mathfrak{f}}' }{\mathfrak{f}} \Big) -\frac{\beta}{\Delta_{r}} \Big( {\alpha}' + \frac{\beta}{\Delta_{r}^{2}} V(r) \Big) , \nonumber \\
U & = \frac{\Delta_{r}^{2}}{\beta} \Big((2 \alpha + \frac{{\beta }'}{\Delta_{r}} - \frac{\beta}{\Delta_{r}} \frac{ {\mathfrak{f}}' }{\mathfrak{f}} )' - \frac{{\gamma}'}{\gamma}  ( \alpha + \frac{{\beta }'}{\Delta_{r}} - \frac{\beta}{\Delta_{r}} \frac{ {\mathfrak{f}}' }{\mathfrak{f}} )\Big) + V(r) , \nonumber \\
G & = \frac{\Delta_{r}}{r^{3}} + \frac{\Delta_{r} {\mathfrak{f} }'}{2 r^{2} \mathfrak{f} } - \frac{{\Delta_{r}}'}{r^{2}} .
\end{align}
The asymptotic solution of $X^{in}_{\ell}$ can be expressed as follows:
\begin{equation}
X_{\ell}^{\mathrm{in}}(r)=\left\{\begin{array}{ll}
A_{\ell}^{trans} e^{-i \omega r^{*}} & r^{*} \rightarrow-\infty, \\
A_{\ell}^{out} e^{i \omega r^{*}}+A_{\ell}^{in} e^{-i \omega r^{*}} & r^{*} \rightarrow+\infty.
\end{array}\right.\label{XAA}
\end{equation}
Meanwhile, the inverse transformation is described by the following expression:
\begin{align}
	 R(r) = & \frac{1}{\gamma} \left[ \Big( \alpha + \frac{{\beta }'}{\Delta_{r}} - \frac{\beta}{\Delta_{r}} \frac{ {\mathfrak{f}}' }{\mathfrak{f}} \Big) \frac{\Delta_{r}}{r \mathfrak{f}^{1/2}} X(r) - \frac{\beta}{\Delta_{r}} (\frac{\Delta_{r}}{r \mathfrak{f}^{1/2}} X(r))' \right] . \label{XRtransform}
\end{align}
Using a method similar to that used in Refs. \cite{Sasaki_2003,Poisson_1997}, the coefficient $A_{\ell}^{in}$ in \eqref{XAA} is related to $B_{\ell}^{in}$ in Eq. (\ref{Kk}) as follows:
\begin{eqnarray}\label{AinBin}
B^{in}_{\ell}=
 -\frac{1}{4\omega^2}A^{in}_{\ell}.
\end{eqnarray}

\subsubsection{Solution of $X_{\ell}^{in} $ }
We now look for the solution of $X^{in}_{\ell}$ and amplitude $A^{in}_{\ell}$ of the S-N-like equation. The method employed in this subsection is based on the work of Sasaki \cite{Sasaki_2003} and Mino \cite{Mino_1997}. We first take the following ansatz:
\begin{equation}\label{Xxi}
X_{\ell}^{in} = e^{- i \phi(z)} z \xi_{\ell}(z),
\end{equation}
where $z=\omega r$, $\eta=2 G M \omega$, $b_{1} =  \frac{c_{1}^{3}}{(c_{1} - c_{2})(c_{1} - c_{h})}$, $b_{2} =  \frac{c_{2}^{3}}{(c_{2} - c_{1})(c_{2} - c_{h})}$, $b_{h} =  \frac{c_{h}^{3}}{(c_{1} - c_{h})(c_{h} - c_{2})}$, and
\begin{align}
	\nonumber \phi(z) & = \int ( \frac{r^{2} \omega}{\Delta_{r}} - \omega ) \mathrm{d} r \\
	& = \eta (b_{1} \ln (z - c_{1} \eta) + b_{2} \ln (z - c_{2} \eta) - b_{h} \ln (z - c_{h} \eta)).
\end{align}
With this choice of the phase function, $\xi_{\ell m}$ is regular and finite at $z = \eta c_{h}$. Then, we determine that Eq.~\eqref{eqX} can be expressed as follows:
\begin{equation}\label{LL}
L^{(0)} \left[ \xi_{\ell} \right] = \eta L^{(1)} \left[ \xi_{\ell} \right] + \eta^{2} L^{(2)} \left[ \xi_{\ell} \right] + \eta^{3} L^{(3)} \left[ \xi_{\ell} \right] + \mathcal{O}(\eta^{4}) ,
\end{equation}
with
\begin{align}
	L^{(0)} = & \frac{\mathrm{d}^{2}}{\mathrm{d} z^{2}} + \frac{2}{z} \frac{\mathrm{d}}{\mathrm{d} z} + \left[ 1 - \frac{\ell (\ell + 1)}{z^{2}} \right] , \nonumber \\
	L^{(1)} = & \frac{1}{z} \frac{\mathrm{d}^{2}}{\mathrm{d} z^{2}} + \frac{1 + \frac{2}{3} a_{2} + 2 i z}{z^{2}} \frac{\mathrm{d}}{\mathrm{d} z} - \frac{ 4 + z^{2} + \frac{a_{2} (\ell^{2} + \ell + 2)}{3} - i z}{z^{3}}  , \nonumber \\
	L^{(2)} = & - \frac{a_{2}}{4 z^{2}} \frac{\mathrm{d}^{2}}{\mathrm{d} z^{2}} + \Big( \frac{i}{z^{2}} \mathfrak{a}_{\ell}^{(2)}  + \frac{\mathfrak{b}_{\ell}^{(2)}}{z^{3}} \Big) \frac{\mathrm{d}}{\mathrm{d} z} + \Big( \frac{\mathfrak{c}_{\ell}^{(2)}}{z^{2}} + \frac{i}{z^{3}} \mathfrak{d}_{\ell}^{(2)} + \frac{\mathfrak{e}_{\ell}^{(2)}}{z^{4}} \Big) , \nonumber \\
    L^{(3)} = & - \frac{a_{3}}{8 z^{3}} \frac{\mathrm{d}^{2}}{\mathrm{d} z^{2}} + \Big( \frac{\mathfrak{a}_{\ell}^{(3)}}{z^{2}} + \frac{i}{z^{3}} \mathfrak{b}_{\ell}^{(3)} + \frac{\mathfrak{c}_{\ell}^{(3)}}{z^{4}} \Big) \frac{\mathrm{d}}{\mathrm{d} z} + \Big( \frac{i}{z^{2}} \mathfrak{d}_{\ell}^{(3)} + \frac{\mathfrak{e}_{\ell}^{(3)}}{z^{3}} + \frac{i}{z^{4}} \mathfrak{f}_{\ell}^{(3)} + \frac{\mathfrak{g}_{\ell}^{(3)}}{z^{5}} \Big) , \label{eq:the expression of 3 order differential operators}
\end{align}
where the definitions of $\mathfrak{a}_{\ell}^{(n)}$ and other terms are shown in Appendix \ref{ap: definitionsThe definitions of hick}.

In the low-frequency limit and noting that $\eta = 2 G M \omega$ only appears on the right-hand side of Eq.~\eqref{LL}, we may look for the solution of the $\xi_{\ell }(z)$ perturbative in terms of $\eta$, i.e.,
\begin{eqnarray}
\xi_{\ell }(z) =\sum_{n=0}^{\infty}\eta^n \xi^{(n)}_{\ell  }(z),
\end{eqnarray}
and we obtain the recursive equation from Eq.~\eqref{LL} as follows:
\begin{eqnarray}
L^{(0)}[\xi^{(n)}_{\ell }]=W^{(n)}_{\ell },
\end{eqnarray}
where
\begin{eqnarray}
W^{(0)}_{\ell }&=&0 \label{epsilon0}, \\
W^{(n)}_{\ell }&=& \sum_{i = 1}^{n} L^{(i)}[\xi^{(n - i)}_{\ell }] , \quad n = 1, 2, 3.
\end{eqnarray}

The solution of $\xi_{\ell}^{(0)}$ can be expressed as a linear combination of the spherical Bessel functions $j_{\ell}$ and $n_{\ell}$, i.e., $\xi_{\ell}^{(0)}=\alpha^{(0)}  j_{\ell} + \beta^{(0)}  n_{\ell} $. Because $n_{\ell} $ does not match with the horizon solution at the leading order of $\eta$, we should take $\beta^{(0)}=0$. Furthermore, because the constant $\alpha^{(0)}$ represents the overall normalization of the solution, which can be chosen arbitrarily, we set $\alpha^{(0)}=1$. That is, for the zeroth-order solution, we have $f_{\ell}^{(0)} = j_{\ell}$ and $g_{\ell}^{(0)} = 0$. Then, one can immediately write the integral expression for $\xi_{\ell}^{(n)}$ (where $n > 0$). Noting that $j_{\ell} {n_{\ell}}' - n_{\ell} {j_{\ell}}' = 1/z^{2}$, we derive the expression of $\xi_{\ell}^{(\beta)}$ for $\beta \geq 1$ as follows:
\begin{align}
	\xi_{\ell}^{(\beta)} = & n_{\ell} \int^{z} \mathrm{d} z \Big(z^{2} j_{\ell} W_{\ell}^{(\beta)} \Big) - j_{\ell} \int^{z} \mathrm{d} z \Big( z^{2} n_{\ell} W_{\ell}^{(\beta)} \Big) . \label{eq:Using the solution of homogeneous equation to construct the solution of inhomogeneous equation}
\end{align}

In general, $\xi_{\ell}^{(\beta)}$ can be decomposed into the real and imaginary parts $\xi_{\ell}^{(\beta)} = f_{\ell}^{(\beta)} + i g_{\ell}^{(\beta)}$, in which
\begin{align}
	& f_{\ell}^{(\beta)} = n_{\ell} \int^{z} \mathrm{d} z \Big(z^{2} j_{\ell} \mathbf{Re} [W_{\ell}^{(\beta)}] \Big) - j_{\ell} \int^{z} \mathrm{d} z \Big( z^{2} n_{\ell} \mathbf{Re} [W_{\ell}^{(\beta)}] \Big) , \label{eq:thefl} \\
	& g_{\ell}^{(\beta)} = n_{\ell} \int^{z} \mathrm{d} z \Big(z^{2} j_{\ell} \mathbf{Im} [W_{\ell}^{(\beta)}] \Big) - j_{\ell} \int^{z} \mathrm{d} z \Big( z^{2} n_{\ell} \mathbf{Im} [W_{\ell}^{(\beta)}] \Big) ,
\end{align}
with
\begin{align}	
	& \mathbf{Re} [W_{\ell}^{(\beta)}] = \sum_{i = 1}^{\beta} \left( \mathbf{Re}[L^{(i)}] [f^{(\beta - i)}_{\ell }] - \mathbf{Im}[L^{(i)}] [g^{(\beta - i)}_{\ell }] \right) , \\
	& \mathbf{Im} [W_{\ell}^{(\beta)}] = \sum_{i = 1}^{\beta} \left( \mathbf{Im}[L^{(i)}] [f^{(\beta - i)}_{\ell }] + \mathbf{Re}[L^{(i)}] [g^{(\beta - i)}_{\ell }] \right) ,
\end{align}
where $\mathbf{Re}[x]$ and $\mathbf{Im}[x]$ are the real and imaginary parts of $x$, respectively. Using the previously presented formula and the method discussed in Appendix \ref{ap:general}, after some tedious calculations, we derive closed analytical formulas of the ingoing-wave S-N-like function for arbitrary $\ell$ to the first order of $\eta$. At the second order of $\eta^{2}$, we can calculate results for any order utilizing equation \eqref{eq:thefl}. However, generalizing these results to encompass all values of $\ell $ is unattainable. Therefore, for the higher-order results, we only provide results for specific values of $\ell$, for $\ell =2, 3$ to $\eta^{2}$ order, and for $\ell = 2$ to $\eta^{3}$ order. We express the real parts as follows:
\begin{align}
	\nonumber f_{\ell}^{(1)} = & \frac{(\ell - 1)(\ell + 3)}{2 (\ell + 1) (2 \ell + 1)} j_{\ell + 1} - \left(\frac{\ell^{2} - 4}{2 \ell (2 \ell + 1)} + \frac{2 \ell - 1}{\ell(\ell-1)}\right) j_{\ell-1} + R_{\ell, 0} j_{0}  - 2 D_{\ell}^{n j} + \\
	& \sum_{m=1}^{\ell-2} \left(\frac{1}{m} + \frac{1}{m+1} \right) R_{\ell, m} j_{m} - \frac{a_{2}}{3} \Big( \frac{\ell^{2} + 3 \ell + 4}{2 (\ell + 1) (2 \ell + 1)} j_{\ell + 1} - \frac{\ell^{2} - \ell + 2}{2 \ell (2 \ell + 1)} j_{\ell - 1} \Big) ,
\end{align}
where $R_{m, k}$ is the Lommel polynomial ($R_{m,k} = - R_{k,m}$ for $m < k$), expressed as follows:
\begin{align}
	\nonumber R_{m,k} & = z^{2} (n_{m} j_{k} - j_{m} n_{k}) \quad (m > k) \\
	& = - \sum_{r = 0}^{[ \frac{(m - k - 1)}{2} ]}  \frac{(-1)^{r} (m - k - 1 - r)! \Gamma \left( m + \frac{1}{2} - r \right)}{r! (m - k - 1 - 2 r)! \Gamma \left( k + \frac{3}{2} + r \right)} \left( \frac{2}{z} \right)^{m - k - 1 - 2 r},
\end{align}
$B_{J}$ is the generalized integral sinusoidal function, and $D_{\ell}^{J}$ is the generalized spherical Bessel function in Appendix \ref{ap:general}.
\begin{align}
	%2
	\nonumber f_{2}^{(2)} = & \left( - \frac{193 a_{2}^2}{1890 z} + \frac{45 a_{2}}{14 z} - \frac{257 a_{3}}{1008 z} - \frac{113}{420 z} \right) j_{1} + \left( - \frac{17 a_{2}^2}{1890 z} - \frac{10 a_{2}}{63 z} - \frac{59 a_{3}}{336 z} + \frac{1}{7 z} \right) j_{3} + \\
	\nonumber & \left( -\frac{16 a_{2}^2}{945}+\frac{a_{2}}{21}-\frac{5 a_{3}}{126}-\frac{107}{210} \right) j_{2} \ln z  + \left( \frac{32 a_{2}^2}{315 z}-\frac{55 a_{2}}{21 z}+\frac{5 a_{3}}{21 z} \right) n_{0} + \\
	\nonumber & \left( \frac{32 a_{2}^2}{945}-\frac{2 a_{2}}{21}+\frac{5 a_{3}}{63}+\frac{107}{105} \right) D_{-3}^{nj} + \left( \frac{10}{3}-\frac{2 a_{2}}{15} \right) D_{1}^{nj} + \frac{14 a_{2}}{45}  D_{3}^{nj} - \frac{389 j_{0}}{70 z^2} - \\
	& \frac{1}{2} j_{2} (\ln z)^{2} + \frac{6 D_{0}^{nj}}{z}-\frac{5 D_{2}^{nj}}{3 z}+4 D_{2}^{nnj} , \\
	%3
	\nonumber f_{3}^{(2)} = & \left( \frac{20 a_{2}^2}{81 z}-\frac{635 a_{2}}{72 z}+\frac{5 a_{3}}{9 z}-\frac{445}{14 z^3}-\frac{1031}{588 z} \right) j_{0} + \left( -\frac{197 a_{2}^2}{1134 z}+\frac{1093 a_{2}}{168 z}-\frac{415 a_{3}}{1008 z}+\frac{323}{49 z} \right) j_{2} \\
	\nonumber & + \left(\frac{2 a_{2}^2}{189 z}-\frac{199 a_{2}}{840 z}-\frac{65 a_{3}}{504 z}+\frac{1}{4 z}\right) j_{4} + j_{3} \ln z \left( -\frac{5 a_{2}^2}{378}+\frac{a_{2}}{42}-\frac{5 a_{3}}{168}-\frac{13}{42} \right) + 4 D_{3}^{nnj} + \\
	\nonumber & \left( \frac{25 a_{2}^2}{63 z}-\frac{405 a_{2}}{28 z}+\frac{25 a_{3}}{28 z}-\frac{65}{6 z} \right) n_{1} +\left(-\frac{5 a_{2}^2}{189}+\frac{a_{2}}{21}-\frac{5 a_{3}}{84}-\frac{13}{21}\right) D_{-4}^{nj} + \left(\frac{13}{3}-\frac{8 a_{2}}{63}\right) D_{2}^{nj} \\
	& + \frac{11 a_{2}}{42} D_{4}^{nj} - \frac{5065 j_{1}}{294 z^2} - \frac{1}{2} j_{3} (\ln z)^{2} + \frac{65 n_{0}}{6 z^2}+\frac{30 D_{0}^{nj}}{z^2}+\frac{9 D_{1}^{nj}}{z}-\frac{3 D_{3}^{nj}}{z} .
\end{align}
\begin{align}
\nonumber f_{2}^{(3)} = & \left( \frac{9}{4} - \frac{a_{2}}{30} \right) j_{1} (\ln z)^{2} + \left( \frac{7 a_{2}}{90} - \frac{1}{12} \right) j_{3} (\ln z)^{2} + D_{2}^{nj} (\ln z)^{2} + \Big( - \frac{16 a_{2}^3}{14175} + \frac{5 a_{2}^2}{63} - \frac{a_{3} a_{2}}{378} \\
\nonumber & - \frac{887 a_{2}}{3150} + \frac{5 a_{3}}{28} + \frac{349}{140} \Big) j_{1} \ln z + \frac{ c_{1}^{2} + ( c_{2} - 1 ) ( c_{1} + c_{2} ) }{z} j_{2} \ln z + \left( \frac{2}{3} - \frac{28 a_{2}}{45} \right) D_{3}^{nnj} + \\
\nonumber & \left( \frac{16 a_{2}^3}{6075} - \frac{11 a_{2}^2}{315} + \frac{a_{3} a_{2}}{162} + \frac{1543 a_{2}}{18900} - \frac{10 a_{3}}{189} + \frac{29}{252} \right) j_{3} \ln z + \Big( -\frac{16 a_{2}^2}{315} + \frac{a_{2}}{7} - \frac{5 a_{3}}{42} - \\
\nonumber & \frac{107}{70} \Big) n_{0} \ln z + \left( \frac{32 a_{2}^2}{945} - \frac{2 a_{2}}{21} + \frac{5 a_{3}}{63} + \frac{107}{105} \right) D_{2}^{nj} \ln z + \Big( - \frac{2381 a_{2}^3}{66150} + \frac{8207 a_{2}^2}{132300} - \\
\nonumber & \frac{185 a_{3} a_{2}}{2352} - \frac{83821 a_{2}}{44100} + \frac{21}{100} - \frac{187 a_{3}}{168} \Big) j_{1} + \Big( \frac{97 a_{2}^3}{36450} + \frac{5339 a_{2}^2}{170100} + \frac{35 a_{3} a_{2}}{3888} + \frac{9053 a_{2}}{25200} + \\
\nonumber & \frac{4609 a_{3}}{18144} - \frac{457}{1050} \Big) j_{3} + \left( \! -\frac{11 a_{2}^3}{142884} + \frac{139 a_{2}^2}{476280} - \frac{43 a_{3} a_{2}}{95256} - \frac{277 a_{2}}{105840} + \frac{1}{504} - \frac{109 a_{3}}{45360} \right) j_{5} \\
\nonumber & + \left( \frac{193 a_{2}^3}{10206} - \frac{110 a_{2}^2}{1701} + \frac{1033 a_{3} a_{2}}{27216} + \frac{6911 a_{2}}{5670} + \frac{48353 a_{3}}{90720} - \frac{3 (\ln z)^{2}}{2} - \frac{2539}{3780} \right) n_{0} + \\
\nonumber & \left( - \frac{32 a_{2}^3}{1215} + \frac{65 a_{2}^2}{1134} - \frac{5 a_{3} a_{2}}{81} - \frac{1574 a_{2}}{945} + \frac{197}{126} - \frac{2455 a_{3}}{3024} \right) n_{2} + \Big( \frac{32 a_{2}^3}{6075} - \frac{58 a_{2}^2}{2835} + \\
\nonumber & \frac{a_{3} a_{2}}{81} + \frac{824 a_{2}}{4725} - \frac{5 a_{3}}{378} - \frac{107}{630} \Big) D_{-4}^{nj} + \Big( \! - \frac{32 a_{2}^3}{14175} - \frac{2 a_{2}^2}{45} - \frac{a_{3} a_{2}}{189} + \frac{7468 a_{2}}{1575} - \frac{5 a_{3}}{42} - \\
\nonumber & \frac{457}{70} \Big) D_{-2}^{nj} + \left( -\frac{19 a_{2}^2}{567} + \frac{59 a_{2}}{21} - \frac{2629}{630} - \frac{103 a_{3}}{1512} \right) D_{0}^{nj} + \Big( \frac{1402 a_{2}^2}{19845} - \frac{925 a_{2}}{441} + \frac{1165 a_{3}}{5292} + \\
\nonumber & \frac{16949}{4410} \Big) D_{2}^{nj} + \left( \frac{17 a_{2}^2}{6615} + \frac{20 a_{2}}{441} + \frac{59 a_{3}}{1176} - \frac{2}{49} \right) D_{4}^{nj} + \left( - \frac{64 a_{2}^2}{945} + \frac{4 a_{2}}{21} - \frac{10 a_{3}}{63} - \frac{214}{105} \right) D_{2}^{njj} + \\
& \left( \! - \frac{64 a_{2}^2}{945} + \frac{4 a_{2}}{21} - \frac{10 a_{3}}{63} - \frac{214}{105} \! \right) D_{-3}^{nnj} - 12 D_{-1}^{nnj} + \left( \frac{4 a_{2}}{15} - 18 \right) D_{1}^{nnj} - 8 D_{2}^{nnnj} .
\end{align}
The corresponding imaginary parts are expressed as follows:
\begin{align}
	g_{\ell}^{(1)} = & j_{\ell} \ln z , \label{eq:imagine g1} \\
	g_{\ell}^{(2)} = & f_{\ell}^{(1)} \ln z - \frac{\mathbb{T}_{1}}{z} j_{\ell} + \varsigma_{\ell}^{(2)} , \label{eq:imagine g2} \\
	g_{\ell}^{(3)} = & f_{\ell}^{(2)} \ln z - \frac{\mathbb{T}_{1} f_{\ell}^{(1)}}{z}  -  \frac{\mathbb{T}_{2} j_{\ell}}{2 z^{2}} + \frac{1}{3} j_{\ell} (\ln z)^{3} + \varsigma_{\ell}^{(3)} , \label{eq:imagine g3}
\end{align}
with
\begin{align}	
	\mathbb{T}_{1} = & 1 + c_{1}^{2} - (c_{1} + c_{2})(1 - c_{2}), \nonumber \\ \mathbb{T}_{2} = & c_{1} c_{2} (-c_{1}-c_{2}+3)+2 (c_{1}-1) c_{1}+2 (c_{2}-1) c_{2} + 1 ,
\end{align}
where $\varsigma_{\ell}^{(n)}$ for $\ell = 2$ can be expressed as follows:
\begin{align}
	%2
	\varsigma_{2}^{(2)} = & \left( \frac{2 a_{2}^2}{81} + \frac{5 a_{3}}{108} + \frac{a_{2}}{180} \right) j_{3} + \frac{a_{2}}{30} j_{1} , \\
	\nonumber \varsigma_{2}^{(3)} = & \left(-\frac{4 a_{2}^2}{81}-\frac{a_{2}}{90}-\frac{5 a_{3}}{54}\right) D_{3}^{nj} +  \! \left( \! \frac{176 a_{2}^3}{2835 z}-\frac{1469 a_{2}^2}{3780 z}+\frac{40 a_{2} a_{3}}{189 z}-\frac{5 a_{2}}{24 z}-\frac{181 a_{3}}{252 z} \! \right) \! j_{1} - \frac{a_{2}}{15} D_{1}^{nj}  \\
	\nonumber & +\! \left( \frac{5 a_{2}^3}{1701 z} \! + \frac{83 a_{2}^2}{3780 z} \! + \frac{20 a_{2} a_{3}}{567 z} \! + \! \frac{a_{2}}{144 z} \! + \! \frac{13 a_{3}}{252 z} \! \right) j_{3} + \left( \frac{22 a_{2}^3}{2835}+\frac{a_{2}^2}{105}+\frac{5 a_{2} a_{3}}{189}+\frac{a_{3}}{56} \right) j_{2} \ln z \\
	& \! + \! \left( \! - \! \frac{44 a_{2}^3}{945 z} \! + \! \frac{296 a_{2}^2}{945 z} \! - \! \frac{10 a_{2} a_{3}}{63 z} \! + \! \frac{a_{2}}{12 z} \! + \! \frac{37 a_{3}}{63 z} \! \right) \! n_{0} \! + \! \left( \! - \! \frac{44 a_{2}^3}{2835} \! - \! \frac{2 a_{2}^2}{105} \! - \! \frac{10 a_{2} a_{3}}{189} \! - \! \frac{a_{3}}{28} \! \right) \! D_{\!- \!3}^{nj} .
\end{align}

Inserting these expressions into Eq.~\eqref{Xxi} and expanding the result with respect to $\eta$, we find that
\begin{align}\label{eq:the solution of X}
	X_{\ell m \omega}^{in} = X_{\ell}^{(0)} +\eta X_{\ell}^{(1)}+\eta^2 X_{\ell}^{(2)} +\eta^3 X_{\ell}^{(3)},
\end{align}
where
\begin{align}
	X_{\ell}^{(0)} = & z j_{\ell} , \nonumber \\
	 X_{\ell}^{(1)} =& z f_{\ell}^{(1)} ,  \nonumber \\
	X_{\ell}^{(2)} = & z \Big[ f_{\ell}^{(2)} + \frac{1}{2} j_{\ell} (\ln z)^{2} + i \varsigma_{\ell}^{(2)} \Big] ,  \nonumber \\
	X_{\ell}^{(3)} = & z \Big[ f_{\ell}^{(3)} + \frac{1}{2} f_{\ell}^{(1)} (\ln z)^{2} - \frac{\mathbb{T}_{1}}{z} j_{\ell} \ln z + \varsigma_{\ell}^{(2)} \ln z + i \varsigma_{\ell}^{(3)} \Big] .
\end{align}

\subsubsection{Coefficient of amplitude $A_{\ell}^{in}$}
Noting $e^{- i \eta (- b_{1} \ln (z - c_{1} \eta ) + b_{2} \ln (z - c_{2} \eta ) + b_{h} \ln (z - c_{h} \eta ))}  = e^{- i z^{*}} e^{i z} \stackrel{z \rightarrow \infty}{\longrightarrow} 1$, taking the expressions of the spherical Hankel functions of the first and second kinds $h_{\ell}^{(1)}$ and $h_{\ell}^{(2)}$ as
\begin{align}
h_{\ell}^{(1)} & = j_{\ell} + i n_{\ell} \xrightarrow{z \rightarrow \infty} (- i)^{\ell + 1} \frac{e^{i z}}{z} , \\
h_{\ell}^{(2)} & = j_{\ell} - i n_{\ell} \xrightarrow{z \rightarrow \infty} i^{\ell + 1} \frac{e^{- i z}}{z},
\end{align}
and using the asymptotic behavior of $B_{J}$ and $D_{\ell}^{J}$ in Ref.~\cite{Mino_1997}, we obtain the following expression:
\begin{align}\label{eq:amplitude}
\nonumber A_{\ell}^{in} = & \frac{1}{2} i^{\ell+1} e^{-i \eta  (\ln 2 \eta + \mathbf{elg} )} e^{i \big[\eta  p_{\ell}^{(0)} -\pi  \eta ^2 p_{\ell}^{(1)} + \eta ^3 \big(p_{\ell}^{(2)}-\pi ^2 p_{\ell}^{(3)}+ p_{\ell}^{(4)} \mathbf{RiZ}(3) \big)\big]} \bigg\{ 1 - \frac{\pi}{2} \eta + \\
& \eta ^2 \Big[ 2 ( \mathbf{elg} +\ln 2) p_{\ell}^{(1)}+q_{\ell}^{(1)}+\frac{5 \pi ^2}{24} \Big] + \eta ^3 \Big[ \pi  q_{\ell}^{(2)}+\pi ^3 q_{\ell}^{(4)}+\pi  ( \mathbf{elg} +\ln 2) q_{\ell}^{(3)} \Big] \bigg\}.
\end{align}
where $\mathbf{elg}$ is the EulerGamma constant ($\mathbf{elg} = 0.57721 \cdots$), $\mathbf{RiZ}(n)$ is the Riemann zeta function ($\mathbf{RiZ}(3) = 1.202 \cdots$), and the coefficients of $A_{2}^{in}$ are
\begin{align}
p_{2}^{(0)} = & \frac{15-2 a_{2}}{9} , \quad p_{2}^{(1)} = \frac{32 a_{2}^2-90 a_{2}+75 a_{3}+963}{3780} , \nonumber \\
p_{2}^{(2)} = & \frac{-292 a_{2}^3+1962 a_{2}^2-765 a_{2} a_{3}-1782 a_{2}+3564 a_{3}+1566}{34992} , \nonumber \\
p_{2}^{(3)} = & \frac{32 a_{2}^2-90 a_{2}+75 a_{3}+963}{11340} , \quad p_{2}^{(4)} = \frac{1}{3} , \nonumber \\
q_{2}^{(1)} = & \frac{-37 a_{2}-5 a_{3}+150}{108} , \quad q_{2}^{(4)} = - \frac{1}{16}, \nonumber \\ q_{2}^{(2)} = & \frac{-176 a_{2}^3-216 a_{2}^2-600 a_{2} a_{3}+7770 a_{2}+645 a_{3}-31500}{45360} , and \nonumber  \\
q_{2}^{(3)} = & \frac{-32 a_{2}^2+90 a_{2}-75 a_{3}-963}{3780} - i \frac{176 a_{2}^3+216 a_{2}^2+600 a_{2} a_{3}+405 a_{3}}{22680 \pi } .
\end{align}

\subsection{Quasi-circular orbit on the equatorial plane around an EOB} \label{subsec:quasi-circularorbit}
In this section, we consider a quasi-circular orbit. In this case, we assume that the orbit lies on the equatorial plane ($\theta = \pi/2$) without loss of generality. By setting $V_{r}(r_{0}) = \partial V_{r}/\partial r (r_{0}) = 0$, the effective energy $E$ and effective angular momentum $L$ are given by
\begin{align}
E/\mu = & \frac{ \sqrt{2} \big[a_{3} (G M)^{3} + r_{0} \big(a_{2} (G M)^{2} + r_{0} (r_{0} - 2 G M)\big)\big] }{ r_{0} \sqrt{r_{0}} \sqrt{5 a_{3} (G M)^{3} + 2 r_{0} \big(2 a_{2} (G M)^{2} + r_{0} (r_{0} - 3 G M) \big)} } , \\ \label{eq:theenergy}
L/\mu = & \frac{r_{0} \sqrt{- 3 a_{3} (G M)^{3} + 2 G M r_{0} (r_{0} - a_{2} G M)}}{\sqrt{5 a_{3} (G M)^{3} + 2 r_{0} \big(2 a_{2} (G M)^{2} + r_{0} (r_{0} - 3 G M) \big)}} .
\end{align}
where $r_{0}$ is the orbital radius.
By defining
\begin{align}
&{}_{0}b_{\ell m} =  \frac{\sqrt{(\ell - 1) \ell (\ell + 1) (\ell + 2)}}{2 \sqrt{2 \pi} \Delta_{r} } \frac{15 a_{3} (G M)^{2} + 8 a_{2} G M r_{0} - 6 r_{0}^{2}}{5 a_{3} (G M)^{2} + 4 a_{2} G M r_{0} - 6 r_{0}^{2}} {}_{0}Y_{\ell m}(\frac{\pi}{2},0) \times r_{0}^{2} E , \nonumber  \\
&{}_{- 1}b_{\ell m} =  \frac{2 \sqrt{(\ell - 1) (\ell + 2)}}{\sqrt{2 \pi} r_{0}} \frac{5 a_{3} (G M)^{2} + 3 a_{2} G M r_{0} - 3 r_{0}^{2}}{\Big(5 a_{3} (G M)^{2} + 4 a_{2} G M r_{0} - 6 r_{0}^{2}\Big)^{3}} {}_{- 1}Y_{\ell m}(\frac{\pi}{2},0) \cdot \mathcal{P}_{r} \cdot L ,\nonumber   \\
&{}_{- 2}b_{\ell m} =  \frac{\Delta_{r} }{\sqrt{2 \pi} r^{4} E} {}_{- 1}Y_{\ell m}(\frac{\pi}{2},0) L^{2} ,\nonumber   \\
&\mathcal{P}_{r} = 75 a_{3}^{2} (G M)^{4} + 10 a_{3} (G M)^{2} r_{0} \big( 11 a_{2} G M - 18 r_{0} \big) + 4 r_{0}^{2} \big(8 a_{2}^2 (G M)^2 - 21 a_{2} G M r_{0} + 9 r_{0}^2\big),\nonumber \\
&\mathcal{B}_{r} = r_{0} \Delta_{r} \Big( 8 a_{2} r_{0} + 30 a_{3} G M \Big) + 6 a_{3} r_{0}^{2} (G M)^{2} - 8 a_{2} a_{3} r_{0} (G M)^{3} - 15 a_{3}^{2} (G M)^{4},
\end{align}
we obtain
\begin{align}
\nonumber \mathbf{A}_{0} = & \frac{1}{2 r_{0}^{2}} \cdot \Big\{ 2 {}_{0}b_{\ell m} + 4 i {}_{-1}b_{\ell m} \Big[ 1 + \frac{i}{2} \omega \frac{r_{0}^{3}}{\Delta_{r} } \Big( 1 + \frac{2 G M r_{0} }{\mathcal{P}_{r}} \times \\
\nonumber & \big( 6 a_{2} r_{0}^{2} + 30 a_{3} G M r_{0} - 5 a_{2} a_{3} (G M)^{2} \big) \Big) \Big] - 2 i \frac{{}_{- 2}b_{\ell m} \omega r_{0}^{3}}{\Delta_{r}^{2} } \Big[ r_{0}^{2} - \\
\nonumber & G M r_{0} + \frac{G M \times \mathcal{B}_{r}}{15 a_{3} (G M)^{2} + 8 a_{2} G M r_{0} - 6 r_{0}^{2}} + \frac{1}{2} i \omega r_{0}^{3} + \\
& 6 i \frac{G M \Delta_{r}^{2}  \cdot \Big( 2 a_{2} r_{0}^{2} + 15 a_{3} G M r_{0} - 5 a_{2} a_{3} (G M)^{2} \Big)}{r_{0}^{2} \omega \big( 15 a_{3} (G M)^{2} + 8 a_{2} G M r_{0} - 6 r_{0}^{2} \big)^{2}} \Big] \Big\} , \\
\nonumber \mathbf{A}_{1} = & i \frac{{}_{-1}b_{\ell m}}{r_{0}} (1 + \frac{2 G M r_{0} \Big( 6 a_{2} r_{0}^{2} + 30 a_{3} G M r_{0} - 5 a_{2} a_{3} (G M)^{2} \Big)}{\mathcal{P}_{r}}) - \\
& \frac{{}_{- 2}b_{\ell m}}{r_{0}} (1 + \frac{G M \cdot (15 a_{3} G M + 4 a_{2} r_{0})}{15 a_{3} (G M)^{2} + 8 a_{2} G M r_{0} - 6 r_{0}^{2}} + i \frac{r_{0}^{3} \omega}{\Delta_{r} }) , \\
\mathbf{A}_{2} = & - \frac{{}_{- 2}b_{\ell m}}{2} . \label{eq:A2}
\end{align}

\section{Energy radiation rate and gravitational waveforms} \label{sec:energy}
Inserting the aforementioned result of $\mathbf{A}_{i} (i = 1,2,3)$, $B_{\ell}^{in}$, and $R_{\ell}^{in}$ into Eq.~\eqref{ZZSch}, we can obtain the following expression:
\begin{align}
Z_{\ell m \omega_{0}} = & Z_{\ell m \omega_{0}}^{(N,\epsilon)} \tilde{Z}_{\ell m \omega_{0}} , \\
Z_{\ell m \omega_{0}}^{(N,\epsilon)} = & (- 1)^{\ell + \epsilon + 1} m^2 \frac{\nu}{2 M} n_{\ell m}^{\left(\epsilon\right)} v^{\ell + \epsilon + 6} {}_{0}Y_{\ell-\epsilon,-m}(\pi / 2, \varphi) , \\
n_{\ell m}^{(\epsilon)} = & \begin{cases}
(i m)^{\ell} \frac{8 \pi}{(2 \ell + 1)!!} \sqrt{\frac{(\ell + 1)(\ell + 2)}{\ell (\ell - 1)}} , & \epsilon = 0 , \\
- (i m)^{\ell} \frac{16 \pi}{(2 \ell + 1)!!} \sqrt{\frac{(2 \ell + 1)(\ell + 2)(\ell^{2} - m^{2})}{(2 \ell - 1)(\ell + 1)\ell (\ell - 1)}} , & \epsilon = 1 .
\end{cases}
\end{align}
where we define $v = (G M \Omega)^{1/3}$, $\omega_{0} = m \Omega$, $\epsilon = 1$ when $\ell + m = 1$, and $\epsilon = 0$ when $\ell + m = 0$. We can divide the higher-order term $\tilde{Z}_{\ell m \omega_{0}}$ into two parts: the $\tilde{Z}_{\ell m \omega_{0}}^{(S)}$ is computed in the Schwarzschild case \cite{panyi_PhysRevD.83.064003} and the $\tilde{Z}_{\ell m \omega_{0}}^{(R)}$ is the 2PM and 3PM perturbation terms:
\begin{equation}
	\tilde{Z}_{\ell m \omega_{0}} = \tilde{Z}_{\ell m \omega_{0}}^{(S)} + \tilde{Z}_{\ell m \omega_{0}}^{(R)} .
\end{equation}
The explicit expression of $\tilde{Z}_{\ell m \omega_{0}}^{(R)}$ is presented in Appendix \ref{ap:thez}. In the test particle limit, i.e., $\nu \rightarrow 0$, we note that $\tilde{Z}_{\ell m \omega_{0}}^{(R)}$ vanishes completely because $a_{2}$ and $a_{3}$ approach $0$. That is, our results revert to the Schwarzschild case in the test particle limit.

\begin{figure}[http]
	\raggedright
	\begin{subfigure}{0.45\textwidth}
		\includegraphics[width=1.1\linewidth]{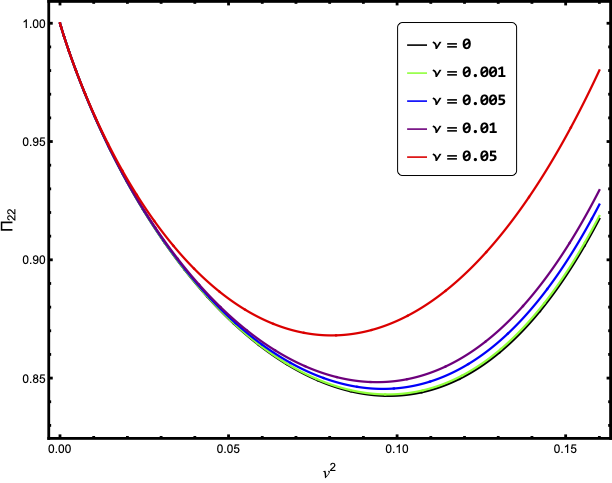}
				\label{fig:sub1}
	\end{subfigure}
	\hspace{2em} % Adjust the vertical space between subfigures
	\begin{subfigure}{0.45\textwidth}
		\includegraphics[width=1.1\linewidth]{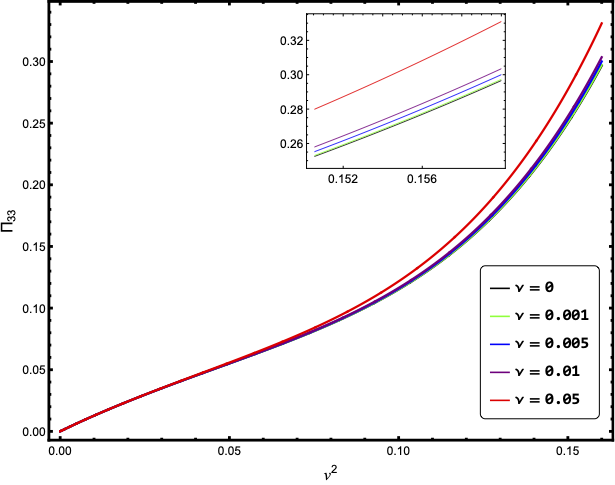}
		\label{fig:sub2}
	\end{subfigure}
	\caption{$\Pi_{\ell m} = \Pi_{\ell m}(v^{2}, \nu)$ with respect to $v^{2}$ and $\nu$, where the curve of $\nu=0$ corresponds to the 4.5PN result obtained for the Schwarzschild case. The figure on the left is for $\Pi_{2 2} = \Pi_{2 2}(v^{2}, \nu)$, and the right one is for $\Pi_{3 3} = \Pi_{3 3}(v^{2}, \nu)$}
	\label{fig:overall}
\end{figure}

In Eq. \eqref{psi41}, utilizing the symmetry of the spin-weighted spherical harmonics, $\,_{s}Y_{\ell,- m}(\frac{\pi}{2},0)= (-1)^{s+\ell} \,_{s}Y_{\ell m}(\frac{\pi}{2},0)$, we know that $Z_{\ell (-m)\omega}=(-1)^{\ell}Z^*_{\ell m \omega}$, where $Z^*_{\ell m \omega}$ is the complex conjugate of $Z_{\ell m \omega}$. In terms of the amplitude $Z_{\ell m\omega}$, we find from Eq.~\eqref{psi41} that the gravitational waveform \cite{Poisson_1993_physrevd.47.1497,Sasaki_1994_10.1143,Buonanno_2007_physrevd.75.124018} at infinity is given by
\begin{align}
& h_{+}-ih_{\times} = \sum_{\ell m} h_{\ell m} \frac{{}_{-2}Y_{\ell m} }{ \sqrt{2\pi}}
	e^{i\omega_{0}(r^*-t)+im\varphi} , \label{waveforms}
\end{align}
with
\begin{align}
& h_{\ell m} = - \frac{2}{\mathcal{R} \omega_{0}^2} Z_{\ell m\omega_{0}} = h_{\ell m}^{(S)} + h_{\ell m}^{(R)} ,
\end{align}
where
\begin{align}
& h_{\ell m}^{(S)} = h_{\ell m}^{(N,\epsilon)} \tilde{Z}_{\ell m \omega_{0}}^{(S)} , \quad h_{\ell m}^{(R)} = h_{\ell m}^{(N,\epsilon)} \tilde{Z}_{\ell m \omega_{0}}^{(R)} , \\
& h_{\ell m}^{(N,\epsilon)} = \frac{G M \nu}{\mathcal{R}} n_{\ell m}^{\left(\epsilon\right)} c_{\ell+\epsilon}(\nu) v^{\ell + \epsilon} {}_{0}Y_{\ell-\epsilon,-m}(\pi / 2, \varphi) .
\end{align}
The energy loss rate along any orbit, in polar coordinates, can be expressed as $\frac{d  E[g_{\mu\nu}^{\text{eff}}]}{d t}= \dot{R} {\cal F}_{R} [g_{\mu\nu}^{\text{eff}}]+\dot{\varphi} {\cal F}_{\varphi}[g_{\mu\nu}^{\text{eff}}]$. By simply replacing the radial component with zero, an excellent approximation of the radiation reaction forces can be obtained \cite{Buonanno_2000}. Thus, from Eq.~\eqref{waveforms}, we know that, for given energy $\omega_n$, the energy loss rate \cite{Poisson_1993_physrevd.47.1497,Buonanno_2007_physrevd.75.124018} for the “plus” and “cross” modes of the gravitational wave is described by the following expression:
\begin{align} \label{dE1}
\frac{dE[g_{\mu\nu}^{\text{eff}}]}{dt} = \frac{1}{2} \left(\frac{\mathrm{d} E}{\mathrm{d} t}\right)_{N} \sum_{\scriptstyle \ell=2}^{\infty} \sum_{m=1}^\ell \Pi_{\ell m},
\end{align}
with
\begin{align}
\Pi_{\ell m} = & \mathcal{E}_{\ell m}^{(S)} + \mathcal{E}_{\ell m}^{(R)}  , \\
\mathcal{E}_{\ell m}^{(S)} = & \frac{ \mid Z_{\ell m \omega_{0} }^{(N,\epsilon)} \tilde{Z}_{\ell m \omega_{0}}^{(S)}\mid^2}{2 \pi G^{2} \omega_{0}^{2} \left(\frac{\mathrm{d} E}{\mathrm{d} t}\right)_{N}} , \\
\mathcal{E}_{\ell m}^{(R)} = & \frac{ \mid Z_{\ell m \omega_{0} }^{(N,\epsilon)} \mid^2}{2 \pi G^{2} \omega_{0}^{2} \left(\frac{\mathrm{d} E}{\mathrm{d} t}\right)_{N}} \Big\{ \tilde{Z}_{\ell m \omega_{0}}^{(S)} \tilde{Z}_{\ell m \omega_{0}}^{(R) *} + \tilde{Z}_{\ell m \omega_{0}}^{(R)} \tilde{Z}_{\ell m \omega_{0}}^{(S) *} + \tilde{Z}_{\ell m \omega_{0}}^{(R)} \tilde{Z}_{\ell m \omega_{0}}^{(R) *} \Big\} .
\end{align}
where $\left(\mathrm{d} E/\mathrm{d} t\right)_{N} = 32 \nu^{2} v^{10}/5$ is the Newtonian quadrupole luminosity and the superscript $*$ denotes the complex conjugation of the corresponding expression. In Figure \ref{fig:overall}, we present the curves of $\Pi_{22}$ and $\Pi_{33}$ as the symmetric mass ratio $\nu$ takes different values.

Then, we determine the radiation reaction forces for the “plus” and “cross” modes of the gravitational wave as follows:
\begin{eqnarray}\label{dEt1}
	{\cal F}_{\varphi}^{\rm circ}[g_{\mu\nu}^{\text{eff}}] \simeq \frac{1}{\dot{\varphi}}\frac{d  E[g_{\mu\nu}^{\text{eff}}]}{d t} = \frac{1}{2} \left(\frac{\mathrm{d} E}{\mathrm{d} t}\right)_{N} \sum_{\scriptstyle \ell=2}^{\infty}\sum_{m=1}^\ell \Big(\mathbf{F}_{\ell m}^{(S)} + \mathbf{F}_{\ell m}^{(R)} \Big),
\end{eqnarray}
where $\mathbf{F}_{\ell m}^{(S)} = \mathcal{E}_{\ell m}^{(S)}/\frac{1}{\dot{\varphi}}$ and $\mathbf{F}_{\ell m}^{(R)} = \mathcal{E}_{\ell m}^{(R)}/\frac{1}{\dot{\varphi}}$.

Eqs.\eqref{waveforms}, \eqref{dE1}, and \eqref{dEt1} indicate that all of the gravitational waveforms, the energy radiation rate, and the radiation reaction forces are based on the effective spacetime.

\section{Conclusions}
In this paper, we investigate the waveforms and energy radiation rate of gravitational waves generated by coalescing spinless binary systems up to 3PM approximation in the EOB theory. We focus on the radiation reaction forces in the Hamilton equation, which can be described by the energy radiation rate $\frac{\mathrm{d} E}{\mathrm{d} t} = \frac{1}{4 \pi G \omega^{2}} \int |\varPsi_{4}^{B}|^{2} r^{2} \mathrm{d} \Omega$ and the “plus” and “cross” modes of gravitational waves, which are related to the null tetrad components of the gravitational perturbed Weyl tensor by $\varPsi_{4}^{B} = \frac{1}{2} (\ddot{h}_{+} - i \ddot{h}_{\times})$. Clearly, to find the energy radiation rate and construct gravitational waveforms, the key step is to seek the solution of $\varPsi_{4}^{B}$. Therefore, the main task of this paper is to solve the decoupled and separated equations of the null tetrad components of gravitational perturbed Weyl tensor $\varPsi_{4}^{B}$ in the effective spacetime by employing the Green's function method.

To achieve this goal, noting that the potential function in the radial Teukolsky-like equation is a long-range potential, we first transform it into an S-N-like equation, which has a short-range potential. Then, by expanding the homogeneous S-N-like equation with $\eta = 2 G M \omega$, where $\omega$ denotes the angular frequency of the wave, we derive closed analytical expressions for the solutions of each order. These solutions are essential for constructing the Green function and asymptotic amplitude. The lowest-order solution is expressed as a linear combination of spherical Bessel functions, which allows us to perform iterative calculations to obtain higher-order solutions. This approach simplifies the problem and enables us to efficiently study the radial Teukolsky-like equation. In the calculation process, we use a low-frequency approximation and considered the conditions of quasi-circular orbits. These conditions are represented by the relationships $z \propto v$, $\eta \propto v^{3}$ and $\omega=m \Omega$. As a result, the obtained results are accurate to $\mathcal{O}(v^{9- 2(\ell - 2) - \epsilon})$, which means that the accuracy of the results of this paper reaches the 4.5PN order\cite{PhysRevD.83.064003,Damour_2009_10.1103/physrevd.79.064004}.
\\\hspace*{1em}
This work also presents a more general integral formula than that given by Sasaki \cite{Mino_1997}, which can be extended to higher orders or even arbitrary orders without additional treatments. In Appendix \ref{ap:general}, the general integral formulas, which can theoretically derive the series solution of the homogeneous S-N-like equation to any order, are presented. However, when constructing the general solution of the nonhomogeneous equation using Green’s function method, it is necessary to specify the amplitude at infinity, which requires finding the asymptotic behavior of $B_{J}$ as $z \to \infty$. Although we know what needs to be done at each step, we have not yet been able to implement our ideas using a computer, but we can obtain specific results through complex calculations. Therefore, combining the outstanding works of Sasaki et al. with the useful formulas presented in Appendix \ref{ap:general}, we have confidence that this method will yield good results in the future.

From the analysis provided, it is evident that the effective metric degenerates into the Schwarzschild case in the test particle limit ($\nu\to 0$). This limit is characterized by vanishing of the coefficients $a_2$ and $a_3$. Therefore, the gravitational waveforms and energy radiation rate calculated in this study were divided into two parts: the Schwarzschild part and the correction part related to PM parameters $a_2$ and $a_3$. Handling spinning binary systems will involve additional complexities. However, the results of this paper and the listed mathematical techniques will be valuable for understanding the energy flux and waveforms in spinning binary systems.

\acknowledgments
%|--------------------------------------------------------------------|
{ We would like to thank professors S. Chen and Q. Pan  for useful discussions on the manuscript. This work was supported by the Grant of Natural Science Foundation of China No. 12035005, and the National Key Research and Development  Program of China No. 2020YFC2201400.}  %|--------------------------------------------------------------------|
\endacknowledgments

\newpage
\section{Appendix}

\begin{appendices}
	
%\newaliascnt{myequation}{equation} % 创建公式的别名
%\aliascntresetthe{myequation} % 重置公式别名的格式
%\newcommand{\myequationautorefname}{Equation} % 设置公式别名的引用前缀

\subsection{The definition of coefficient of $L^{(n)}$} \label{ap: definitionsThe definitions of hick} \label{ap:the definition of hick} %%%%%%
\setcounter{equation}{0} % 将公式别名计数器归为0
\renewcommand{\theequation}{A.\arabic{equation}}
Associated with $L^{(2)}$ are expressed as follows
\begin{align}
\mathfrak{a}_{\ell}^{(2)} = & - \frac{36 a_{2} + 8 \lambda a_{2}^{2} + 3 \lambda \Big[ 5 a_{3} - 6 (\lambda + 2) \big(c_{1}^{2} + (c_{1} + c_{2}) (c_{2} - 1)\big) \Big]}{9 \lambda (\lambda + 2)} , \\
\mathfrak{b}_{\ell}^{(2)} = & \frac{ 4 a_{2}^{2} \lambda (\lambda + 4) + 5 a_{3} \lambda (2 \lambda + 7) - 4 a_{2} (\lambda^{2} + 2 \lambda + 6) }{6 \lambda (\lambda + 2)} , \\
\mathfrak{c}_{\ell}^{(2)} = & \frac{3 \lambda a_{2}^{2} + 60 \lambda a_{3} + 9 a_{2} (\lambda^{2} + 2 \lambda + 16)}{36 \lambda (\lambda + 2)} , \\
\mathfrak{d}_{\ell}^{(2)} = & \frac{8 a_{2}^{2} \lambda (\lambda + 4) + 15 a_{3} \lambda (\lambda + 4) - 24 a_{2} (\lambda^{2} + 2 \lambda + 12)}{36 \lambda (\lambda + 2)} , \\
\mathfrak{e}_{\ell}^{(2)} = & \frac{- 8 a_{2}^{2} \lambda (\lambda^{2} + 19 \lambda + 22) - 15 a_{3} \lambda (2 \lambda^{2} + 27 \lambda + 34) + 12 a_{2} (23 \lambda^{2} + 34 \lambda -48)}{72 \lambda (\lambda + 2)}  .
\end{align}

Related with $L^{(3)}$ are 
\begin{align}
\mathfrak{a}_{\ell}^{(3)} = & - \frac{2 (\lambda a_{2} - 9) (36 a_{2} + 8 a_{2}^{2} \lambda + 15 a_{3} \lambda)}{27 \lambda^{2} (2 + \lambda)^{2}} \\
\nonumber \mathfrak{b}_{\ell}^{(3)} = & - \frac{1}{18 \lambda^{2} (\lambda + 2)^{2}} \Big\{ 16 a_{2}^{3} \lambda (2 \lambda + 5) - 8 a_{2}^{2} \lambda (2 \lambda^{2} - 5 \lambda + 6) - 3 a_{2} \Big[ 3 \lambda^{4} + \\
\nonumber & (12 - 35 a_{3}) \lambda^{3} + (36 - 80 a_{3}) \lambda^{2} + 48 \lambda - 144 \Big] + 3 \lambda \Big[ - a_{3} \Big( 10 \lambda^{2} - 55 \lambda - 60 \Big) + \\
& 12 (c_{1} - 1) (c_{2} - 1) (c_{1} + c_{2}) \lambda (\lambda + 2)^{2}  \Big] \Big\} , \\
\nonumber \mathfrak{c}_{\ell}^{(3)} = & \frac{1}{216 \lambda (\lambda + 2)} \Big\{ 16 a_{2}^{3} \lambda (7 \lambda + 62) - 9 a_{3} \lambda (37 \lambda + 224) - 36 a_{2}^{2} \big(3 \lambda^{2} + 26 \lambda - 24\big) + \\
& 90 a_{2} \big(24 + a_{3} \lambda (5 \lambda + 38)\big) \Big\} , \\
\mathfrak{d}_{\ell}^{(3)} = & - \frac{2 (\lambda a_{2} - 9) (36 a_{2} + 8 a_{2}^{2} \lambda + 15 a_{3} \lambda)}{27 \lambda^{2} (\lambda + 2)^{2}} , \\
\nonumber \mathfrak{e}_{\ell}^{(3)} = & \frac{1}{216 \lambda^{2} (\lambda + 2)^{2}} \Big\{ - 96 a_{2}^{2} \lambda^{2} (2 \lambda - 5) + 64 a_{2}^{3} \lambda^{2} (6 \lambda + 13) + 12 a_{2} \Big[ 9 \lambda^{4} - 144 \lambda + 864 + \\
\nonumber & 3(35a_{3} + 12) \lambda^{3} + 4(55 a_{3} - 9) \lambda^{2} \Big] + 9 \lambda \Big[ a_{3} (3 \lambda^{3} - 28 \lambda^{2} + 232 \lambda + 480) + \\
& 48 \big(c_{1}^{2} + (c_{2} - 1)(c_{1} + c_{2}) \big) \lambda (\lambda + 2)^{2}  \Big] \Big\} , \\
\nonumber \mathfrak{f}_{\ell}^{(3)} = & \frac{1}{108 \lambda^{2} (\lambda + 2)^{2}} \Bigg\{ 8 a_{2}^{3} \lambda^{2} \left( 3 \lambda^{2} + 44 \lambda + 64 \right) - 24 a_{2}^{2} \lambda  \big(3 \lambda ^3 + 23 \lambda ^2 + 34 \lambda - 48 \big) + \\
\nonumber & 3 a_{2} \Big[ 3 \lambda ^4 \left(10 a_{3} - 8 (c_{1}^{2} + (c_{2} - 1)(c_{1} + c_{2}) ) - 3\right) + 576 \lambda - 1728 + \lambda ^3 \Big( 370 a_{3} - 12 \big( 8 (c_{1}^{2} + \\
\nonumber & (c_{2} - 1)(c_{1} + c_{2}) ) + 3 \big) \Big) + \lambda ^2 \Big(560 a_{3} - 12 \big( 8 (c_{1}^{2} + (c_{2} - 1)(c_{1} + c_{2}) ) - 39\big) \Big) \Big] \\
& - 9 \lambda  \Big[ 5 a_{3} \left(4 \lambda ^3+27 \lambda ^2+80 \lambda +48\right) - 12 (c_{1}-1) (c_{2}-1) (c_{1} + c_{2}) \lambda  (\lambda +2)^2  \Big] \Bigg\} , \\ 
\nonumber \mathfrak{g}_{\ell}^{(3)} = & \frac{1}{216 \lambda (\lambda + 2)} \Big\{ -8 a_{2}^3 \lambda  \left( \lambda ^2 + 62 \lambda + 48 \right) + 12 a_{2}^2 \left( 31 \lambda ^2 + 2 \lambda + 192 \right) - \\
& 3 a_{2} \Big( 15 a_{3} \lambda ^3+650 a_{3} \lambda ^2+8 (65 a_{3}-18) \lambda -1440 \Big) + 72 a_{3} \lambda  (23 \lambda +16) \Big\}.
\end{align}
where $\lambda = (\ell - 1) (\ell + 2)$.

\subsection{How to calculate the $\xi_{\ell}^{(n)}$}  \label{ap:general}%%%%%%
\setcounter{equation}{0} % 将公式别名计数器归为0
\renewcommand{\theequation}{B.\arabic{equation}}

\begin{figure}[http]
	\centering
	\includegraphics[scale=0.42]{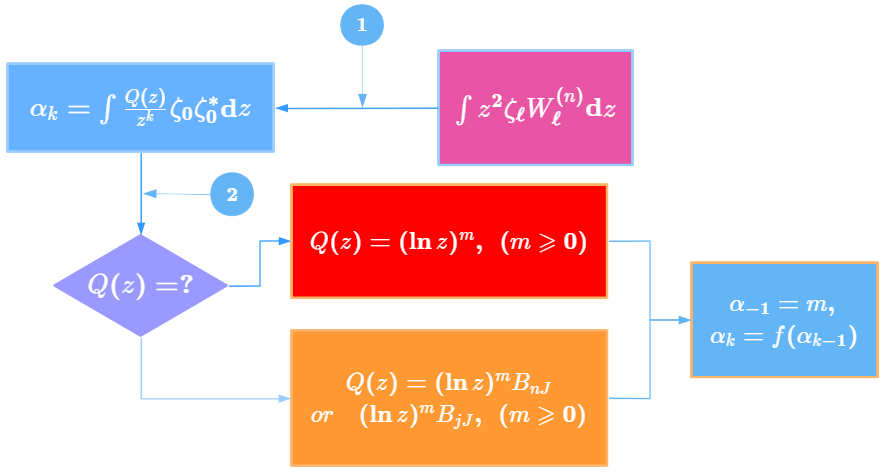}
	\caption{The figure shows how to calculate the integral in Eq.~\eqref{eq:Using the solution of homogeneous equation to construct the solution of inhomogeneous equation}. $\zeta_{i}$ and $\zeta_{i}^{*}$ stands for $j_{i}$ or $n_{i}$, $i$ is an arbitrary integer. \circledtext{tianlan}{1} involves transforming the integrand using the properties of spherical Bessel functions, while \circledtext{tianlan}{2} yields different integral expressions depending on the type of $Q(z)$.}
	\label{fig:howtoev}
\end{figure}

The integral appearing in Eq.~\eqref{eq:Using the solution of homogeneous equation to construct the solution of inhomogeneous equation} involves spherical Bessel functions with different quantum numbers. By utilizing the following formula, 
\begin{align}
& \frac{2 \ell + 1}{z} \zeta_{\ell} = \zeta_{\ell - 1} + \zeta_{\ell + 1} , \quad \zeta \text{ is } j \text{ or } n ,  \\
& n_{\ell} = (- 1)^{\ell + 1} j_{- \ell - 1} \label{item:The relationship between j and n ,1} , \\
& j_{\ell} = (- 1)^{\ell} n_{- \ell - 1} \label{item:The relationship between j and n ,2} , \\
& j_{0} = \frac{\sin z}{z}, \quad n_{0} = - \frac{\cos z}{z} .
\end{align}
we can express any quantum number $\ell$ of \(j_{\ell}\) and \(n_{\ell}\) in the form of \(\textbf{poly1}(z)j_{0} + \textbf{poly2}(z)n_{0}\). \(\textbf{poly1}(z)\) and \(\textbf{poly2}(z)\) are both polynomials in \(1/z\), while \(j_{0}\) and \(n_{0}\) have elementary function representations. This transformation allows for easier computation of the integral expression. Based on our research in mathematical manuals and insights from Sasaki $et$ $al.$\cite{Mino_1997}, we have classified these integrals and derived formal integral formulas, as depicted in Fig.~\ref{fig:howtoev}, outlining the specific implementation route.

\subsubsection{Generalized sine integral function $B_{J}$ and generalized spherical Bessel function $D_{\ell}^{J}$}
We first introduce some special integral formulas from Ref.~\cite{Mino_1997}, the generalized sine integral function $B_{J}$
\begin{align}
	B_{j J} & = \int_{z_{*}}^{z} z j_{0} D_{0}^{J} \mathrm{d} z , \label{eq:BjJ} \\
	B_{n J} & = \int_{z_{*}}^{z} z n_{0} D_{0}^{J} \mathrm{d} z . \label{eq:BnJ}
\end{align}
And the generalized spherical Bessel functions $D_{\ell}^{J}$:
\begin{align}
	D_{\ell}^{j} & = j_{\ell} , \quad D_{\ell}^{n} = n_{\ell} , \\
	D_{\ell}^{n J} & = n_{\ell} B_{j J} - j_{\ell} B_{n J} , \\
	D_{\ell}^{j J} & = j_{\ell} B_{j J} + n_{\ell} B_{n J} .
\end{align}
From the above definitions, we can derive the following expressions when \(J = j, n\):
\begin{align}
	B_{j j} = & \int_{0}^{z} z j_{0} j_{0} \mathrm{d} z = - \frac{1}{2} C(z) , \label{eq:equ of Bjj} \\
	B_{n j} = & \int_{0}^{z} z n_{0} j_{0} \mathrm{d} z = - \frac{1}{2} S(z) , \\
	B_{j n} = & \int_{0}^{z} z j_{0} n_{0} \mathrm{d} z = - \frac{1}{2} S(z) , \\
	B_{n n} = & \int_{z_{*}}^{z} z n_{0} n_{0} \mathrm{d} z = - B_{j j} + \ln z ,
\end{align}
where
\begin{align}
	& S(z) = Si(2 z) = \int_{0}^{2 z} \frac{\sin t}{t} \mathrm{d} t \label{eq:the definition of S(z)} \quad (\left | z \right | < \infty) , \\
	& C(z) = - \int_{2 z}^{\infty} \frac{\cos t}{t} \mathrm{d} t - \gamma - \ln 2 z \label{eq:the definition of C(z)} \quad \left | \arg z \right | < \pi . 
\end{align}
$B_{J}$ is essential to note that when \(J\) ends with \(j\), there may be a logarithmic divergence issue when the integrand is combined with \(n_{0}\). This occurs because we use \(n_{0}^{2} = \frac{1}{z^{2}} - j_{0}^{2}\), which results in \(\frac{1}{z^{2}}\) diverging when \(z_{*} = 0\). Therefore, the terms related to \(j_{0}^{2}\) have their lower limit set at \(z_{*} = 0\), while those related to \(\frac{1}{z^{2}}\) are set at \(z_{*} = 1\).

It is worth noting that all strings $J$ ending with the character $n$ in $B_{J}$ can be expressed in terms of those whose $J$ end with $j$, for example,
\begin{align}
B_{jn} = & B_{nj} , \\
B_{jjn} = & - B_{nnj} + B_{nj} \ln z , \\
B_{jnn} = & 2 B_{jjj} + B_{nnj} - B_{jj} \ln z , \\
B_{nnn} = & 2 B_{njj} - B_{jnj} - B_{nj} \ln z .
\end{align}
Using these relations, we can express all the $D_{\ell}^{J}$ whose $J$ end with $n$ in terms of those whose $J$ end with $j$.

\begin{figure}[http]
	\centering
		\includegraphics[scale=0.64]{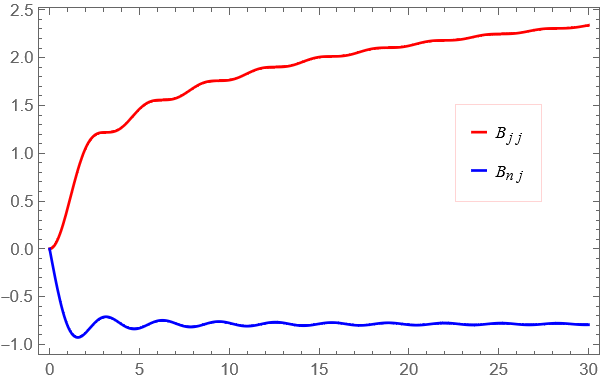}
		\includegraphics[scale=0.64]{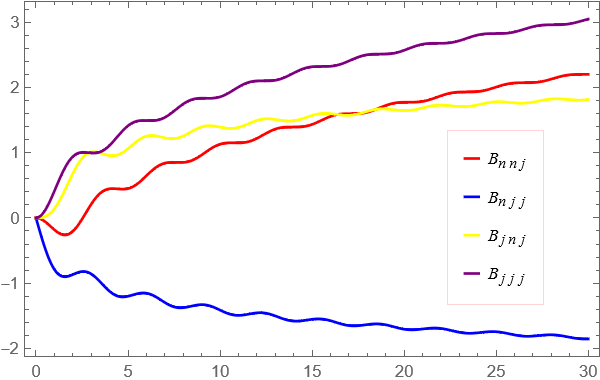}
	\caption{The image of $B_{J}$ for $J$ of length 2 is given on the left, and the image of $B_{J}$ for $J$ of length 3 is given on the right.}
	\label{figureof$B_{J}$}
\end{figure}

\subsubsection{the general integral form of $Q(z)$}
For \(k > -1\) and \(k \in \mathbb{N}\), with $Q(z)$ being differentiable functions, we can establish the following integral properties:
\begin{align}
	& \boxed{\color{red}\int \frac{Q(z)}{z^{k}} j_{0}^{2} \mathrm{d} z} = \frac{2}{1 + k} \Big\{ - \boxed{\color{blue}\int \frac{Q(z)}{z^{k - 1}} n_{0} j_{0} \mathrm{d} z} - \frac{1}{2} \frac{Q(z)}{z^{k - 1}} j_{0}^{2} + \frac{1}{2} \int \frac{{Q(z)}'}{z^{k - 1}} j_{0}^{2} \mathrm{d} z \Big\} , \label{eq:the expression of ABjj} \\ %%%%%
	& \boxed{\color{violet}\int \frac{Q(z)}{z^{k}} n_{0}^{2} \mathrm{d} z} = \frac{2}{1 + k} \Big\{ \boxed{\color{blue}\int \frac{Q(z)}{z^{k - 1}} n_{0} j_{0} \mathrm{d} z} - \frac{1}{2} \frac{Q(z)}{z^{k - 1}} n_{0}^{2} + \frac{1}{2} \int \frac{{Q(z)}'}{z^{k - 1}} n_{0}^{2} \mathrm{d} z \Big\} , \label{eq:the expression of ABnn} \\ %%%%
	& \boxed{\color{blue}\int \frac{Q(z)}{z^{k}} n_{0} j_{0} \mathrm{d} z} = \begin{cases}
		\frac{2}{1 + k} \Big\{ \boxed{\color{red}\int \frac{Q(z)}{z^{k - 1}} j_{0}^{2} \mathrm{d} z} - \frac{1}{2} \int \frac{Q(z)}{z^{k + 1}} \mathrm{d} z - \frac{1}{2} \frac{Q(z)}{z^{k - 1}} n_{0} j_{0} + \frac{1}{2} \int \frac{{Q(z)}'}{z^{k - 1}} n_{0} j_{0} \mathrm{d} z \Big\} , \\
		\frac{2}{1 + k} \Big\{ \frac{1}{2} \int \frac{Q(z)}{z^{k + 1}} \mathrm{d} z - \boxed{\color{violet}\int \frac{Q(z)}{z^{k - 1}} n_{0}^{2} \mathrm{d} z} - \frac{1}{2} \frac{Q(z)}{z^{k - 1}} n_{0} j_{0} + \frac{1}{2} \int \frac{{Q(z)}'}{z^{k - 1}} n_{0} j_{0} \mathrm{d} z \Big\} .
	\end{cases}  \label{eq:the expression of ABnj}
\end{align}
and for \(k = -1\), we define $\int z Q(z) \zeta_{0} \zeta_{0}^{*} \mathrm{d} z$ as a new function. This enables us to establish a complete recursive relationship, allowing us to obtain the desired expressions by letting the computer perform a finite number of iterations for a specific level.

\subsubsection{the case of $Q(z) = (\ln z)^{m}, (m \geqslant 0 , m \in \mathbb{N})$}
Defining \(\mathbb{E}_{k}^{m} = \int \frac{j_{0}^{2} (\ln z)^{m}}{z^{k}} \mathrm{d} z\) and \(\mathbb{F}_{k}^{m} = \int \frac{j_{0} n_{0} (\ln z)^{m}}{z^{k}} \mathrm{d} z\), $m = 0, 1, 2, \dots $ and $k = -1, 0, 1, 2, \dots$, we can derive the following expressions utilizing the integral formulas \eqref{eq:the expression of ABjj}-\eqref{eq:the expression of ABnj} mentioned above:
\begin{align}
	& \begin{tikzcd}[ampersand replacement=\&]
		-1 \arrow[r, green] \& m \\
		k' \arrow[ru, red] \&{}
	\end{tikzcd} 
	\Longrightarrow
	\mathbb{E}_{-1}^{m} , \mathbb{E}_{k'}^{m} \quad k' = 0, 1, 2, \dots \\
	& \begin{tikzcd}[ampersand replacement=\&]
		-1 \arrow[r, green] \& m \\
		0 \arrow[ru, red] \& {} \\
		k'' \arrow[ruu, yellow] \& {}
	\end{tikzcd}
	\Longrightarrow
	\mathbb{F}_{-1}^{m} , \mathbb{F}_{0}^{m} , \mathbb{F}_{k''}^{m} \quad k'' = 1, 2, \dots
\end{align}
where
\begin{align}
	& \mathbb{E}_{-1}^{m} = \mathbf{GenlogS}_{m}^{j} , \quad \mathbb{F}_{-1}^{m} = \mathbf{GenlogU}_{m}^{j} , \\
	& \mathbb{F}_{0}^{m} = 2 \Big\{ \mathbb{E}_{- 1}^{m} - \frac{1}{2 (m + 1)} (\ln z)^{m + 1} - \frac{1}{2} z n_{0} j_{0} (\ln z)^{m} + \frac{m}{2} \mathbb{F}_{0}^{m - 1} \Big\} , \\
	& \mathbb{E}_{k'}^{m} = \frac{2}{1 + k'} \Big\{ - \mathbb{F}_{k' - 1}^{m} - \frac{1}{2} \frac{j_{0}^{2} (\ln z)^{m}}{z^{k' - 1}} + \frac{m}{2} \mathbb{E}_{k'}^{m-1} \Big\} , \\
	& \mathbb{F}_{k''}^{m} = \frac{2}{1 + k''} \Big\{ \mathbb{E}_{k'' - 1}^{m} + \frac{1}{2} \mathbf{ei}_{- m}^{k'' \ln z} (\ln z)^{m + 1} - \frac{1}{2} \frac{n_{0} j_{0} (\ln z)^{m}}{z^{k'' - 1}} + \frac{m}{2}  \mathbb{F}_{k''}^{m - 1} \Big\} ,
\end{align}
$\mathbf{ei}_{n}^{z}$ is an exponential integral of order $n$, \(\mathbf{ei}_{n}^{z} = \int_{1}^{\infty} \frac{e^{- z t}}{t^{n}} \mathrm{d} t\).

\subsubsection{the case of \texorpdfstring{$Q(z) = (\ln z)^{m} B_{jJ}$}{} or \texorpdfstring{$(\ln z)^{m} B_{nJ}, (m \geqslant 0 , m \in \mathbb{N})$}{} }

Defining \(\mathbb{S}_{k}^{m,J} = - \int \frac{j_{0} D_{0}^{nJ} (\ln z)^{m}}{z^{k}} \mathrm{d} z\) and \(\mathbb{U}_{k}^{m,J} = - \int \frac{n_{0} D_{0}^{nJ} (\ln z)^{m}}{z^{k}} \mathrm{d} z\), we can derive the following expressions utilizing the integral formulas mentioned above:
\begin{align}
& \begin{tikzcd}[ampersand replacement=\&]
		-1 \arrow[r, green] \& m \arrow[r, blue] \arrow[rd, blue] \& j \\
		0 \arrow[ru, red] \& {} \& J' \\
		k'' \arrow[ruu, yellow] \& {} \& {} 
	\end{tikzcd} 
	\Longrightarrow \begin{cases}
	& \mathbb{S}_{-1}^{m,J} , \color{purple}\underline{\text{$ \mathbb{S}_{0}^{m,J} , \mathbb{S}_{k''}^{m,J} $}} \\
    & \mathbb{U}_{-1}^{m,J} , \color{purple}\underline{\text{$ \mathbb{U}_{0}^{m,J} , \mathbb{U}_{k''}^{m,J} $}} 
	\end{cases} \Longrightarrow \begin{cases}
	& \Pi_{1}^{m,jJ} , \Pi_{\tilde{k}}^{m,jJ}  \\
	& \Pi_{1}^{m,nJ} , \Pi_{\tilde{k}}^{m,nJ} 
\end{cases}
\end{align}
where $J'$ is a string of length greater than two, and $\tilde{k} = 2, 3, 4, \dots$, the terms $\mathbb{S}_{0}^{m,J} , \mathbb{S}_{k''}^{m,J} $ can actually be merged into a single term $\mathbb{S}_{k'}^{m,J} $. The same holds true for $\mathbb{U}_{0}^{m,J} , \mathbb{U}_{k''}^{m,J} $, as the expressions are entirely determined by $\Pi_{k''}^{m,J}$ due to the parameters $k'$ and $J$. In the following, we provide the corresponding expressions.
\begin{align}
& \mathbb{S}_{-1}^{m,J} = - \mathbf{GenlogS}_{m}^{nJ} , \quad \mathbb{U}_{-1}^{m,J} = - \mathbf{GenlogU}_{m}^{nJ} , \\
& \mathbb{S}_{k'}^{m,J} = \frac{2}{1 + k'} \Big\{ \frac{1}{2} \frac{j_{0} D_{0}^{nJ} (\ln z)^{m}}{z^{k' - 1}} \! - \! \frac{1}{2} \Pi_{k'+1}^{m,jJ} \!-\! \mathbb{U}_{k'-1}^{m,J} \!+\! \frac{m}{2} \mathbb{S}_{k'}^{m-1,J} \Big\} , \\
& \mathbb{U}_{k'}^{m,J} = \frac{2}{1 + k'} \Big\{ \mathbb{S}_{k' - 1}^{m,J} - \frac{1}{2} \Pi_{k' + 1}^{m,nJ} + \frac{1}{2} \frac{n_{0} D_{0}^{nJ} (\ln z)^{m}}{z^{k' - 1}} + \frac{m}{2} \mathbb{U}_{k'}^{m-1,J} \Big\} ,
\end{align}
The definitions of $\Pi_{k'}^{m,jJ}$ and $\Pi_{k'}^{m,nJ}$ are,
\begin{align}
\Pi_{k'}^{m,jJ} = & \int \frac{B_{jJ} (\ln z)^{m}}{z^{k'}} \mathrm{d} z = \begin{cases}
\frac{1}{1 - k'} \Big( \frac{B_{j J} (\ln z)^{m}}{z^{k' - 1}} - m \Pi_{k'}^{m-1,jJ} - \int \frac{j_{0} D_{0}^{J}}{z^{k'-2}} (\ln z)^{m} \mathrm{d} z \Big) , \quad k' > 1 \\
%%%%%
\frac{1}{m + 1} \Big( B_{jJ} (\ln z)^{m+1} - \mathbf{GenlogS}_{m+1}^{J} \Big), \quad k' = 1
\end{cases} \\
\Pi_{k'}^{m,nJ} = & \int \frac{B_{nJ} (\ln z)^{m}}{z^{k'}} \mathrm{d} z = \begin{cases}
	\frac{1}{1 - k'} \Big( \frac{B_{n J} (\ln z)^{m}}{z^{k' - 1}} - m \Pi_{k'}^{m-1,nJ} - \int \frac{n_{0} D_{0}^{J}}{z^{k'-2}} (\ln z)^{m} \mathrm{d} z \Big) , \quad k' > 1 \\
\frac{1}{m + 1} \Big(	B_{nJ} (\ln z)^{m+1} - \mathbf{GenlogU}_{m+1}^{J} \Big) , \quad k' = 1
\end{cases} 
\end{align}
We will then provide the specific expressions of $\Pi_{k'}^{m,jJ}$ and $\Pi_{k'}^{m,nJ}$ for different values of $k'$ and $J$:
\begin{align}
& \Pi_{1}^{m,jj} = \frac{B_{jj} (\ln z)^{m+1} - \mathbf{GenlogS}_{m+1}^{j}}{m + 1}  , \Pi_{1}^{m,nj} = \frac{B_{nj} (\ln z)^{m+1} - \mathbf{GenlogU}_{m+1}^{j}}{m + 1} , \\
& \Pi_{1}^{m,jjJ} = \frac{B_{jjJ} (\ln z)^{m+1} - \mathbf{GenlogS}_{m+1}^{jJ}}{m + 1} , \Pi_{1}^{m,njJ} = \frac{B_{njJ} (\ln z)^{m+1} - \mathbf{GenlogU}_{m+1}^{jJ}}{m + 1} , \\
& \Pi_{1}^{m,jnJ} = \frac{B_{jnJ} (\ln z)^{m+1} - \mathbf{GenlogS}_{m+1}^{nJ}}{m + 1} , \Pi_{1}^{m,nnJ} = \frac{B_{nnJ} (\ln z)^{m+1} - \mathbf{GenlogU}_{m+1}^{nJ}}{m + 1} , \\
& \Pi_{\tilde{k}}^{m,jj} = \frac{1}{1 - \tilde{k}} \Big( \frac{B_{j j} (\ln z)^{m}}{z^{\tilde{k} - 1}} - m \Pi_{\tilde{k}}^{m-1,jj} - \mathbb{E}_{\tilde{k}-2}^{m} \Big) , \\
& \Pi_{\tilde{k}}^{m,nj} = \frac{1}{1 - \tilde{k}} \Big( \frac{B_{nj} (\ln z)^{m}}{z^{\tilde{k} - 1}} - m \Pi_{\tilde{k}}^{m-1,nj} - \mathbb{F}_{\tilde{k}-2}^{m} \Big) , \\
& \Pi_{\tilde{k}}^{m,jjJ} = \frac{1}{1 - \tilde{k}} \Big( \frac{B_{jjJ} (\ln z)^{m}}{z^{\tilde{k} - 1}} - m \Pi_{\tilde{k}}^{m-1,jjJ} - \Pi_{\tilde{k}}^{m,jJ} - \mathbb{U}_{\tilde{k}-2}^{m,J} \Big) , \\
& \Pi_{\tilde{k}}^{m,njJ} = \frac{1}{1 - \tilde{k}} \Big( \frac{B_{njJ} (\ln z)^{m}}{z^{\tilde{k} - 1}} - m \Pi_{\tilde{k}}^{m-1,njJ} - \Pi_{\tilde{k}}^{m,nJ} + \mathbb{S}_{\tilde{k}-2}^{m,J} \Big) , \\
& \Pi_{\tilde{k}}^{m,jnJ} = \frac{1}{1 - \tilde{k}} \Big( \frac{B_{jnJ} (\ln z)^{m}}{z^{\tilde{k} - 1}} - m \Pi_{\tilde{k}}^{m-1,jnJ} + \mathbb{S}_{\tilde{k}-2}^{m,J} \Big) , \\
& \Pi_{\tilde{k}}^{m,nnJ} = \frac{1}{1 - \tilde{k}} \Big( \frac{B_{nnJ} (\ln z)^{m}}{z^{\tilde{k} - 1}} - m \Pi_{\tilde{k}}^{m-1,nnJ} + \mathbb{U}_{\tilde{k}-2}^{m,J} \Big) , 
\end{align}
The definitions of $\mathbf{GenlogS}_{m}^{J}$ and $\mathbf{GenlogU}_{m}^{J}$ are,
\begin{align}
\mathbf{GenlogS}_{m}^{J} = \int z j_{0} D_{0}^{J} (\ln z)^{m} \mathrm{d} z = \sum_{i = 0}^{m} (-1)^{i} \frac{m!}{(m - i)!} (\ln z)^{m - i} \mathbf{theB}_{i + 1}^{jJ} , \\
\mathbf{GenlogU}_{m}^{J} = \int z n_{0} D_{0}^{J} (\ln z)^{m} \mathrm{d} z = \sum_{i = 0}^{m} (-1)^{i} \frac{m!}{(m - i)!} (\ln z)^{m - i} \mathbf{theB}_{i + 1}^{nJ} , 
\end{align}
where
\begin{align}
	& \vartheta_{1}[A] = \{ (1 , A) \} , \\
	& {{ \vartheta_{n}[A] }^{2 \mu - 1} , { \vartheta_{n}[A] }^{2\mu}}=  \begin{cases}
		1. ({\vartheta_{n-1}[A]} ^{\mu 1} , jj{\vartheta_{n-1}[A]} ^{\mu 2} |_{2}^{-1}) , ({\vartheta_{n-1}[A]} ^{\mu 1} , nn{\vartheta_{n-1}[A]} ^{\mu 2} |_{2}^{-1}) , \\
		\quad when \quad {\vartheta_{n-1}[A]} ^{\mu 2} |_{1}^{1} = j , \\
		2. ({\vartheta_{n-1}[A]} ^{\mu 1} , nj{\vartheta_{n-1}[A]} ^{\mu 2} |_{2}^{-1}) ,  (- {\vartheta_{n-1}[A]} ^{\mu 1} , jn{\vartheta_{n-1}[A]} ^{\mu 2} |_{2}^{-1}) , \\
		\quad when \quad {\vartheta_{n-1}[A]} ^{\mu 2} |_{1}^{1} = n .
	\end{cases} \\ 
	& \mathbf{thB}_{i}^{jJ} = { \vartheta_{i}[jJ] }^{\nu 1} B_{{ \vartheta_{i}[jJ] }^{\nu 2}} , \\
	& \mathbf{thB}_{i}^{nJ} = { \vartheta_{i}[nJ] }^{\nu 1} B_{{ \vartheta_{i}[nJ] }^{\nu 2}} , 
\end{align}
in this context, \(\mu = 1, 2, \dots, \text{dim}[\vartheta_{n - 1}[A]]\), \({\vartheta_{n}[A]} ^{i j}\) represents the \((i, j)\)-th element of \({\vartheta_{n}[A]}\), \(\text{dim}[\vartheta_{n}]\) denotes the dimension of \(\vartheta_{n}[A]\), \({\vartheta_{n - 1}[A]} ^{\mu 2} |_{a}^{b}\) refers to the characters within the range from the \(a\)-th to the \(b\)-th in \({\vartheta_{n - 1}[A]} ^{\mu 2}\), and \({\vartheta_{i}[A]} ^{\nu 1} B_{{\vartheta_{i}[A]} ^{\nu 2}}\) implies summation over \(\nu\). We shall provide expressions for the first five terms of \(\mathbf{thB}_{i}^{jJ}\) and \(\mathbf{thB}_{i}^{nJ}\), even though even these initial terms are quite intricate:
\begin{align}
	\mathbf{thB}_{1}^{jJ} = & B_{jJ} , \hspace*{12em} \mathbf{thB}_{1}^{nJ} = B_{nJ} ,  \\
	\mathbf{thB}_{2}^{jJ} = & B_{jjJ} + B_{nnJ} , \hspace*{8.5em} \mathbf{thB}_{2}^{nJ} = B_{njJ} - B_{jnJ} , \\
	\mathbf{thB}_{3}^{jJ} = & B_{jjjJ} + B_{nnjJ} + B_{njnJ} - B_{jnnJ} , \hspace*{0.9em} \mathbf{thB}_{3}^{nJ} = B_{njjJ} - B_{jnjJ} - B_{jjnJ} - B_{nnnJ} , \\
	\mathbf{thB}_{4}^{jJ} = & B_{jjjjJ} + B_{nnjjJ} + B_{njnjJ} - B_{jnnjJ} + B_{njjnJ} - B_{jnjnJ} - B_{jjnnJ} - B_{nnnnJ} , \\
	\mathbf{thB}_{4}^{nJ} = & B_{njjjJ} - B_{jnjjJ} - B_{jjnjJ} - B_{nnnjJ} - B_{jjjnJ} - B_{nnjnJ} - B_{njnnJ} + B_{jnnnJ} , \\
	\nonumber \mathbf{thB}_{5}^{jJ} = & B_{jjjjjJ} + B_{nnjjjJ} + B_{njnjjJ} - B_{jnnjjJ} + B_{njjnjJ} - B_{jnjnjJ} - B_{jjnnjJ} - B_{nnnnjJ} + \\
	& B_{njjjnJ} - B_{jnjjnJ} - B_{jjnjnJ} - B_{nnnjnJ} - B_{jjjnnJ} - B_{nnjnnJ} - B_{njnnnJ} + B_{jnnnnJ} , \\
	\nonumber \mathbf{thB}_{5}^{nJ} = & B_{njjjjJ} - B_{jnjjjJ} - B_{jjnjjJ} - B_{nnnjjJ} - B_{jjjnjJ} - B_{nnjnjJ} - B_{njnnjJ} + B_{jnnnjJ} - \\
	& B_{jjjjnJ} - B_{nnjjnJ} - B_{njnjnJ} + B_{jnnjnJ} - B_{njjnnJ} + B_{jnjnnJ} + B_{jjnnnJ} + B_{nnnnnJ} .
\end{align}

\subsection{the expressions of \texorpdfstring{$\tilde{Z}_{\ell m \omega_{0}}^{(R)}$}{}} 
\label{ap:thez}%%%%%%
\setcounter{equation}{0} % 将公式别名计数器归为0
\renewcommand{\theequation}{C.\arabic{equation}}
We present the expression of $\tilde{Z}_{\ell m \omega_{0}}^{(R)}$ for mode $(\ell, m) = (2, 2), (2, 1), (3, 3)$ in which each mode contains terms up to $\mathcal{O}(v^{9 - 2(\ell - 2) - \epsilon})$.
\begin{align}
\nonumber \tilde{Z}_{2 2 \omega_{0}}^{(R)} &=  a_{2} v^{2} + 2 i a_{2} v^{3} + \frac{308 a_{2}^2-919 a_{2}+420 a_{3}}{378} v^4 + \Big( \frac{41 i a_{2}^2}{18}+a_{2} \big( 12 i \ln v - \frac{238 i}{27} + \\
\nonumber & 4 i \mathbf{elg} + 2 \pi + 12 i \ln 2 \big) + \frac{11 i a_{3}}{6} \Big) v^{5} + \bigg( \frac{44 a_{2}^3}{81} - ( \mathbf{elg} + \ln 2v ) \big( \frac{256 a_{2}^2}{945} + \frac{152 a_{2}}{21} + \\
\nonumber & \frac{40 a_{3}}{63} \big) + \pi  \big( \frac{128 i a_{2}^2}{945} + \frac{76 i a_{2}}{21} + \frac{20 i a_{3}}{63} \big) + \ln 2 \big( -\frac{256 a_{2}^2}{945} - \frac{320 a_{2}}{21} - \frac{40 a_{3}}{63} \big) - \\
\nonumber & \frac{173707 a_{2}^2}{66150} + \frac{85 a_{2} a_{3}}{54} - 16 a_{2} \ln v + \frac{891691 a_{2}}{158760} - \frac{4553 a_{3}}{5292} \bigg) v^{6} + \bigg( \frac{505 i a_{2}^3}{243} - \frac{54365 i a_{2}^2}{6804} \\
\nonumber & + ( \mathbf{elg} + \ln 2 v ) \big( \frac{88 i a_{2}^2}{27} - \frac{1838 i a_{2}}{189} + \frac{40 i a_{3}}{9} \big) + \big( \frac{176 i a_{2}^2}{27} - \frac{3676 i a_{2}}{189} + \frac{80 i a_{3}}{9} \big) \ln 2 v \\
\nonumber & + \pi \big( \frac{44 a_{2}^2}{27}-\frac{919 a_{2}}{189}+\frac{20 a_{3}}{9} \big) + \frac{403 i a_{2} a_{3}}{81}+\frac{67805 i a_{2}}{6804}-\frac{17005 i a_{3}}{2268} \bigg) v^{7} + \bigg( \frac{2 a_{2}^4}{27} - \\
\nonumber & \frac{291169 a_{2}^3}{132300} + \frac{55 a_{2}^2 a_{3}}{162} + \frac{16150588 a_{2}^2}{2083725} + \big( - \frac{256 a_{2}^3}{945} - \frac{513614 a_{2}^2}{19845} - \frac{40 a_{2} a_{3}}{63} + \frac{70328 a_{2}}{735} \\
\nonumber & - \frac{26966 a_{3}}{1323} \big) \ln v + \pi  \big( \frac{128 i a_{2}^3}{945} + \frac{75997 i a_{2}^2}{19845} + \frac{20 i a_{2} a_{3}}{63} - \frac{83236 i a_{2}}{6615} + \frac{3781 i a_{3}}{1323} \big) + \\
\nonumber & \mathbf{elg} \big( -\frac{256 a_{2}^3}{945}-\frac{151994 a_{2}^2}{19845}-\frac{40 a_{2} a_{3}}{63}+\frac{166472 a_{2}}{6615}-\frac{7562 a_{3}}{1323} \big) + \ln 2 \big( - \frac{512 a_{2}^3}{945} - \\
\nonumber & \frac{484798 a_{2}^2}{19845} - \frac{80 a_{2} a_{3}}{63} + \frac{188728 a_{2}}{2205} - \frac{24826 a_{3}}{1323} \big) - \frac{1885 a_{2} a_{3}}{441}-72 a_{2} (\ln 2 v)^2 + \\
\nonumber & 24 i \pi  a_{2} \ln 2 v - 48 \mathbf{elg}  a_{2} \ln 2 v - 8 \mathbf{elg} ^2 a_{2} + \frac{2 \pi ^2 a_{2}}{3}+\frac{2319196003 a_{2}}{366735600}+8 \mathbf{elg}  i \pi  a_{2} + \frac{125 a_{3}^2}{864} + \\
\nonumber & \frac{1009699 a_{3}}{222264} \bigg) v^{8} + \bigg( \frac{352 i a_{2}^4}{243} - \frac{8946247 i a_{2}^3}{1786050} - \big( \frac{1}{315} 1024 i a_{2}^2 + \frac{944 i a_{2}}{7} + \frac{160 i a_{3}}{21} \big) (\ln v)^{2} + \\
\nonumber & \ln 2 \big( -\frac{1}{105} 1024 i a_{2}^2-\frac{1824 i a_{2}}{7}-\frac{160 i a_{3}}{7} \big) \ln v + \pi \ln v \big( -\frac{2048 a_{2}^2}{945}-\frac{880 a_{2}}{21}-\frac{320 a_{3}}{63} \big) + \\
\nonumber & \mathbf{elg} \ln v \big( -\frac{1}{945} 4096 i a_{2}^2-\frac{1760 i a_{2}}{21}-\frac{640 i a_{3}}{63} \big) + \frac{440}{81} i a_{2}^2 a_{3} - \mathbf{elg}^2 \big( \frac{1024 i a_{2}^2}{945} + \frac{272 i a_{2}}{21} + \\
\nonumber & \frac{160 i a_{3}}{63} \big) + \pi^2 \big( \frac{256 i a_{2}^2}{567} + \frac{4 i a_{2}}{63} + \frac{200 i a_{3}}{189} \big) - \mathbf{elg} \pi \big( \frac{1024 a_{2}^2}{945} + \frac{272 a_{2}}{21} + \frac{160 a_{3}}{63} \big) - \\
\nonumber & (\ln 2)^{2} \left( \frac{2048 i a_{2}^2}{315} + \frac{880 i a_{2}}{7} + \frac{320 i a_{3}}{21} \right) - \pi \ln 2 \big( \frac{512 a_{2}^2}{189} + \frac{848 a_{2}}{21} + \frac{400 a_{3}}{63} \big) - \frac{260495 i a_{2} a_{3}}{23814} \\
\nonumber & \mathbf{elg}  \ln 2 \big( \frac{1024 i a_{2}^2}{189} + \frac{1696 i a_{2}}{21} + \frac{800 i a_{3}}{63} \big) + \frac{1183663 i a_{2}^2}{138915} + \big( \frac{17456 i a_{2}^3}{2835} - \frac{2942726 i a_{2}^2}{99225} + \\
\nonumber & \frac{3410 i a_{2} a_{3}}{189} + \frac{716683 i a_{2}}{13230} - \frac{11059 i a_{3}}{1323} - \frac{5136 i}{245} \big) \ln v + \pi \big( \frac{3016 a_{2}^3}{2835} - \frac{339121 a_{2}^2}{99225} + \\
\nonumber & \frac{635 a_{2} a_{3}}{189} + \frac{14639 a_{2}}{15876} + \frac{191 a_{3}}{294} \big) + \mathbf{elg} \big( \frac{6032 i a_{2}^3}{2835} - \frac{678242 i a_{2}^2}{99225} + \frac{1270 i a_{2} a_{3}}{189} + \frac{14639 i a_{2}}{7938} \\
\nonumber & + \frac{191 i a_{3}}{147} \big) + \ln 2 \big( \frac{5776 i a_{2}^3}{945} - \frac{2578726 i a_{2}^2}{99225} + \frac{3490 i a_{2} a_{3}}{189} + \frac{443851 i a_{2}}{13230} - \frac{4787 i a_{3}}{1323} - \\
& \frac{5136 i}{245} \big) + \frac{2503003411 i a_{2}}{91683900} + \frac{175 i a_{3}^2}{108} - \frac{15511 i a_{3}}{16464} \bigg) v^{9} , 
%%%%%%%%%%
%%%%%%%%%%
%%%%%%%%%%
\end{align}

\begin{align}
\nonumber \tilde{Z}_{2 1 \omega_{0}}^{(R)} &=  \frac{a_{2}}{3} v^{2} + \frac{23 i a_{2}}{36} v^{3} + \Big( \frac{5 a_{2}^2}{72}-\frac{20 a_{2}}{63}+\frac{a_{3}}{12} \Big) v^{4} + \bigg( \frac{101 i a_{2}^2}{432} + 2 i a_{2} \ln v + \frac{2 i a_{2} (\mathbf{elg} + \ln 4)}{3} - \\
\nonumber & \frac{37 i a_{2}}{56} + \frac{\pi  a_{2}}{3} + \frac{i a_{3}}{6} \bigg) v^{5} + \bigg( \frac{23 a_{2}^3}{324} + (\mathbf{elg} + \ln 2 v) \big( -\frac{64 a_{2}^2}{945}-\frac{137 a_{2}}{126}-\frac{10 a_{3}}{63} \big) + \\
\nonumber & \pi \big( \frac{32 i a_{2}^2}{945} + \frac{137 i a_{2}}{252} + \frac{5 i a_{3}}{63} \big) + \frac{139673 a_{2}^2}{1587600} + \frac{17 a_{2} a_{3}}{108} - \frac{23}{9} a_{2} \ln v + \frac{3709 a_{2}}{158760} - \\
\nonumber & \frac{23}{18} a_{2} \ln 2 + \frac{2143 a_{3}}{5292} \bigg) v^{6} + \bigg( \frac{199 i a_{2}^3}{2592} + \big( \frac{5 i a_{2}^2}{36} - \frac{40 i a_{2}}{63} + \frac{i a_{3}}{6} \big) \big( \ln 4 v^3 + \mathbf{elg} \big) + \\
\nonumber & \pi \big( \frac{5 a_{2}^2}{72}-\frac{20 a_{2}}{63}+\frac{a_{3}}{12} \big)-\frac{3581 i a_{2}^2}{18144}+\frac{55 i a_{2} a_{3}}{216}+\frac{5291 i a_{2}}{4536}+\frac{2533 i a_{3}}{3024} \bigg) v^{7} + \bigg( -\big( \frac{64 a_{2}^3}{2835} \\
\nonumber & + \frac{10 a_{2} a_{3}}{189} \big) (\ln 2 v + \mathbf{elg}) + \frac{452071 a_{2}^3}{9525600} - \big( \frac{101 a_{2}^2}{216} - \frac{37 a_{2}}{28} + \frac{a_{3}}{3} \big) \ln (2 v^2) + \big( -\frac{6403 a_{2}^2}{17640} + \\
\nonumber & \frac{4643 a_{2}}{8820} - \frac{209 a_{3}}{882} \big) (\ln 2 v + \mathbf{elg} ) + \frac{15354247 a_{2}^2}{133358400} + \pi  \big( \frac{32 i a_{2}^3}{2835} + \frac{6403 i a_{2}^2}{35280} + \frac{5 i a_{2} a_{3}}{189} - \frac{4643 i a_{2}}{17640} \\
\nonumber & + \frac{209 i a_{3}}{1764} \big) + \frac{7037 a_{2} a_{3}}{31752} - \frac{8}{3} a_{2} \ln 2 \big( \ln 2 v^3 + \mathbf{elg} - \frac{i \pi }{2} \big) - 6 a_{2} (\ln v)^{2} - 4 \big( \mathbf{elg} - \frac{i \pi }{2} \big) a_{2} \ln v + \\
& \frac{\pi ^2 a_{2}}{18} - \frac{2 \mathbf{elg} ^2 a_{2}}{3} + \frac{159578843 a_{2}}{122245200} + \frac{2 i \mathbf{elg} \pi a_{2}}{3} + \frac{3985 a_{3}}{296352} \bigg) v^{8} .
\end{align}

\begin{align}
\nonumber \tilde{Z}_{3 3 \omega_{0}}^{(R)} &=  \frac{4 a_{2}}{3} v^{2} + \frac{29 i a_{2}}{12} v^{3} + \Big( \frac{4 a_{2}^2}{3}-5 a_{2} + \frac{3 a_{3}}{2} \Big) v^{4} + \Big( \frac{511 i a_{2}^2}{144} + 8 i a_{2} \big( \ln (12 v^3) + \mathbf{elg} - \frac{i \pi }{2} \big) - \\ \nonumber & \frac{2951 i a_{2}}{120} + \frac{47 i a_{3}}{24} \Big) v^{5} + \bigg( \frac{92 a_{2}^3}{81} - \big( \frac{10 a_{2}^2}{21} + \frac{191 a_{2}}{14} + \frac{15 a_{3}}{14} \big) \big( \ln 6 v + \mathbf{elg} - \frac{i \pi }{2} \big) - \frac{122987 a_{2}^2}{21168} + \\
\nonumber & 
\frac{77 a_{2} a_{3}}{27} - \frac{29}{2} a_{2} \ln 2 v^2 + \frac{571751 a_{2}}{21560} - \frac{16229 a_{3}}{7056} \bigg) v^{6} + \bigg( \frac{1117 i a_{2}^3}{288} - \frac{745 i a_{3}}{32} + \frac{10256 i a_{2}}{165} + \\
& \big( 8 i a_{2}^2 - 30 i a_{2} + 9 i a_{3} \big) \big( \ln (12 v^3) + \mathbf{elg} - \frac{i \pi }{2} \big) - \frac{39121 i a_{2}^2}{1440} + \frac{353 i a_{2} a_{3}}{48} \bigg) v^{7} .
\end{align}

\end{appendices}

\bibliography{reference.bib}

\end{document}